\documentclass[12pt]{article}
\usepackage[xdvi]{graphicx}
\usepackage{amssymb}
\newcommand{\bea}{\begin{eqnarray}}
\newcommand{\eea}{\end{eqnarray}}

\makeatletter
\@addtoreset{equation}{section}
\makeatother

\begin{document}

\begin{titlepage}
\begin{flushright}
OU-HET 651/2009
\end{flushright}

\vspace{25ex}

\begin{center}
{\Large\bf 
Forward-backward asymmetry on Z resonance \\
\vspace{1ex}
in SO(5)$\times$U(1) gauge-Higgs unification
}
\end{center}

\vspace{1ex}

\begin{center}
{\large
Nobuhiro Uekusa
}
\end{center}
\begin{center}
{\it Department of Physics, 
Osaka University \\
Toyonaka, Osaka 560-0043
Japan} \\
\textit{E-mail}: 
uekusa@het.phys.sci.osaka-u.ac.jp
\end{center}


\vspace{3ex}

\begin{abstract}

We find that the tree-level predictions 
of the forward-backward production asymmetries
on the $Z$ resonance for $b$ and $c$ quarks,
$A_{FB}$,
in an $SO(5)\times U(1)$ 
gauge-Higgs unification model
are markedly close to
the central values of the Particle Data Group data
unlike the standard model.
The decay width of $Z$ boson is evaluated
and the $S$ and $T$ parameters are discussed.

\end{abstract}
\end{titlepage}

\newpage

\section{Introduction}

The standard model of 
elementary particles describes physics
well up to the weak scale.
It is a gauge theory with 
quarks and leptons as particles of matter
and with gauge bosons as particles of force.
In addition to matter and force,
the ground state is quantified in a unified framework. 
If the theory were in a symmetric phase, 
the quarks, leptons and gauge bosons 
would be all massless.
To yield their masses,
the gauge symmetry is broken.
A single scalar field minimally represents an 
effective source of gauge symmetry breaking,
though it is unknown even whether further microscopic structure
of the symmetry breaking is possible.
In the standard model, 
quarks, leptons, gauge bosons
and even the Higgs boson are treated on an equal footing.
This is a feasible unified model.
However the standard model has the gauge hierarchy problem.
From an intuitive viewpoint,
while fermions and gauge bosons are called particles of
matter and force, respectively,
it seems obscure what particle the physical Higgs boson
should be called.

The gauge-Higgs unification scenario is 
to treat gauge fields and Higgs field literally
in a unified way.
The four-dimensional 
Higgs field is identified with a part of the extra-dimensional
component of gauge fields in higher 
dimensions~\cite{Fairlie:1979at, Manton:1979kb}.
Due to a physical Wilson line phase,
the electroweak symmetry is dynamically broken~\cite{%
Hosotani:1983xw, Hosotani:1988bm, Davies:1987ei}.
The non-locality of the Wilson line phase yields
a finite mass to the physical Higgs boson and 
it is a solution to the gauge hierarchy problem~\cite{%
Hatanaka:1998yp}.

In a gauge theory with group $SO(5)\times U(1)$,
the custodial symmetry of the standard model
is applied to the gauge-Higgs unification scenario.
It leads to the decisive prediction.
The presence of top quark which is a part of fermions
introduced in the vectorial representation of $SO(5)$
in the bulk five-dimensional Randall-Sundrum spacetime
dynamically induces the electroweak symmetry breaking
where the effective potential for 
the Wilson line phase $\theta_H$ 
is minimized at $\theta_H =\pm {1\over 2}\pi$~\cite{%
Hosotani:2008tx}.
The four-dimensional Higgs field $H(x)$ corresponds to 
four-dimensional fluctuations of the Wilson line phase.
The effective interactions are determined for
definite matter content.
The $WWH$, $ZZH$ and Yukawa couplings 
are suppressed by a factor $\cos \theta_H$
compared with those in the standard model~\cite{%
Giudice:2007fh, 
Hosotani:2008by}.
At $\theta_H = \pm {1\over 2} \pi$ for the potential minimum, 
the couplings with a single Higgs field vanish. 
If the annihilation rate of Higgs bosons is small,
it can become a candidate of the dark matter 
in the universe~\cite{%
Hosotani:2009jk}.
From an intuitive viewpoint, the physical Higgs boson might be
called a particle of dark matter.

Recently, an $SO(5)\times U(1)$ gauge-Higgs unification model
in the Randall-Sundrum spacetime has been generalized to
inclusion of
three generations of fermions (quarks and leptons)
and gauge bosons, i.e., 
all particles of matter and force at the weak scale~\cite{HNU}.
The effective chiral theory of the model is anomaly free.
In the model which holds the distinctive prediction of 
dynamical gauge symmetry breaking
and the stable Higgs boson,
the $W$, $Z$ and electromagnetic currents have been
determined.
The electroweak currents depend on profiles of wave functions
of $W$, $Z$ bosons and quarks and leptons both 
in the fifth-dimension
and $SO(5)$ group.
Despite the highly nontrivial profiles,
it has been found that the deviations of the couplings of fermions
to gauge bosons from the standard model are less than
1\% except for top quark.
This small deviation raises the expectation that
this $SO(5)\times U(1)$ gauge-Higgs unification model 
may be regarded as a 
realistic model.
Once these gauge couplings 
are indicated in an expression of observed quantities,
they can be compared with experimental data 
rather than the values of the standard model.
A representative measured quantity relevant to gauge couplings of
fermions is the forward-backward asymmetry
on the $Z$ resonance, which is denoted as $A_{FB}$.
In the standard model, $A_{FB}$ has been extensively studied
including radiative corrections.
For $b$ quark production, there seems to be
a discrepancy between the standard model prediction and 
the experimental value.
It is certain that 
$A_{FB}$ is a decisive indication to express
the gauge couplings also in the $SO(5)\times U(1)$ model.

In this paper, we present the tree-level 
predictions of
the forward-backward production asymmetries
on the $Z$ resonance for quarks and leptons,
$A_{FB}$,
in the model given in Ref.~\cite{HNU}.
It is found that 
the tree-level prediction for $b$ quark production
gives $A_{FB}^b=0.09952$,
which is quite close to the central value of 
the experimental data
$A_{FB}^b(\textrm{Exp}.)=0.0992\pm 0.0016$.\footnote{%
The experimental value is quoted from the
Particle Data
Group data~\cite{PDG}.}
We find for $c$ quark production  $A_{FB}^c=0.07073$ 
which is also close to the central value of 
the experimental data
$A_{FB}^c(\textrm{Exp}.)=0.0707\pm 0.0035$.
For all fermions, the tree-level predictions of
$A_{FB}$ are given,
and it is shown that the values are not very sensitive 
to whether input parameters for quarks and leptons
correspond to
their running masses at the $m_Z$ scale or 
their pole masses.
Because the gauge-Higgs unification scenario is a
higher-dimensional
theory, it is non-renormalizable and its radiative effects 
should be treated appropriately.
We discuss quantum loop corrections and
divergences for observed quantities.
We also evaluate other electroweak quantities
such as the decay width of $Z$ boson
and the $S$ and $T$ parameters.

The paper is organized as follows.
In Section~\ref{sec:model}, our model is summarized.
In Section~\ref{zcurrent}, we give the equations for
the $Z$ boson couplings of 
fermions required for calculating $A_{FB}$.
The numerical analysis of $A_{FB}$ is 
shown in Section~\ref{afb}. 
The decay width and the $S$ and $T$ parameters
are evaluated in Section~\ref{sec:other}.
Summary and discussions
are given in Section~\ref{conclusion}.
Details of the notation and 
the results for input parameters different than in the main text
are given in Appendices~\ref{app:mode} and \ref{app:wi},
respectively.
The values of the $S$ and $T$ parameters
at one-loop in the standard model are
summarized in Appendix~\ref{app:smst}.

\section{Model \label{sec:model}}

We work on the model given in Ref.~\cite{HNU}.
The model is defined in the Randall-Sundrum (RS) warped spacetime
whose metric is given by~\cite{%
Randall:1999ee, 
Randall:1999vf}
\bea
    ds^2 = G_{MN}dx^M dx^N
  = e^{-2\sigma(y)}\eta_{\mu\nu} dx^\mu dx^\nu + dy^2 ,
      \label{metric1}
\eea
where $\eta_{\mu\nu} =\textrm{diag}(-1,1,1,1)$,
$\sigma(y)=\sigma(y+2L)$, and
$\sigma(y)=k|y|$ for $|y|\leq L$.
The fundamental region in the fifth dimension
is given by $0\leq y\leq L$.
The Planck brane and the TeV brane are located at 
$y=0$
and $y=L$, respectively.  
The bulk region $0 < y < L$ is  an anti-de Sitter spacetime with the 
cosmological constant  $\Lambda = - 6k^2$.

We consider an $SO(5) \times U(1)_X$ gauge
theory in the RS warped spacetime. 
The $SO(5) \times U(1)_X$  symmetry is broken  to $SO(4) \times U(1)_X$
by the orbifold boundary conditions at the Planck and TeV branes.
The symmetry is spontaneously
broken to $SU(2)_L \times U(1)_Y$ by additional
interactions at the Planck brane. 
Here we do not address a question of how 
the orbifold structure of spacetime appears with orbifold conditions.

The action integral consists of four parts:
\bea
  S =
  S_{\textrm{\scriptsize bulk}}^{%
  \textrm{\scriptsize gauge}}
  +S_{\textrm{\scriptsize Pl. brane}}^{%
  \textrm{\scriptsize scalar}}
  +S_{\textrm{\scriptsize bulk}}^{%
  \textrm{\scriptsize fermion}}
  +S_{\textrm{\scriptsize Pl. brane}}^{%
  \textrm{\scriptsize fermions}} .
\eea
The bulk parts respect $SO(5)\times U(1)_X$ gauge
symmetry. There are 
$SO(5)$ gauge fields $A_M$ and $U(1)_X$
gauge field $B_M$. The former are decomposed as 
$A_M = \sum_{I=1}^{10} A_M^I T^I 
    = \sum_{a_L=1}^3 A_M^{a_L} T^{a_L}
     +\sum_{a_R=1}^3 A_M^{a_R} T^{a_R}
     +\sum_{\hat{a}=1}^4 A_M^{\hat{a}} T^{\hat{a}}$,
where  
$T^{a_L,a_R}(a_L,a_R=1,2,3)$ and
$T^{\hat{a}}(\hat{a}=1,\ldots,4)$ 
are the generators of
$SO(4)\sim SU(2)_L \times SU(2)_R$ and $SO(5)/SO(4)$,
respectively.
In a vectorial representation, the components of the 
generator are
$T_{ij}^{a_L,a_R}
   =
     -{i\over 2}
      [{1\over 2} \epsilon^{abc}
       (\delta_i^b \delta_j^c
       -\delta_j^b \delta_i^c
       )
     \pm 
     (\delta_i^a \delta_j^4
     -\delta_j^a \delta_i^4)
    ]$
and 
$T_{ij}^{\hat{a}}
  =-{i\over \sqrt{2}}
   (\delta_i^{\hat{a}}
   \delta_j^5
   -\delta_j^{\hat{a}}
    \delta_i^5)$,
where $i,j=1,\ldots, 5$
and
$\textrm{Tr}(T^I T^J)=\delta^{IJ}$.
The action integral for pure gauge boson part is
\bea
   S_{\textrm{\scriptsize bulk}}^{%
  \textrm{\scriptsize gauge}}
  &\!\!\!=\!\!\!& \int d^5x \sqrt{-G}
  \left[ -\textrm{tr}( {1\over 4} F^{(A)MN} F_{MN}^{(A)}
            +{1\over 2\xi}
  (f_{\textrm{\scriptsize gf}}^{(A)})^2 + 
  {\cal L}_{\textrm{\scriptsize gh}}^{(A)})
  \right.
\nonumber
\\
  &&\qquad \left.
   -({1\over 4} F^{(B)MN}F_{MN}^{(B)} 
  +{1\over 2\xi}
 (f_{\textrm{\scriptsize gf}}^{(B)})^2
    +{\cal L}_{\textrm{\scriptsize gh}}^{(B)})
   \right] ,
\eea
where the gauge fixing and ghost terms are
denoted as functionals with suffixes gf and gh, respectively. Here 
$F_{MN}^{(A)} =
 \partial_M A_N -\partial_N A_M -ig_A  [A_M, A_N]$ 
and 
$F_{MN}^{(B)} =\partial_M B_N -\partial_N B_M$.

The orbifold boundary conditions
at $y_0=0$ and $y_1=L$ for gauge fields are given by
\bea
  && \left(\begin{array}{c}
    A_\mu \\
    A_y \\
   \end{array}\right) (x,y_j-y)
  = P_j \left(\begin{array}{c}
    A_\mu \\
    -A_y \\
   \end{array} \right) (x,y_j+y) P_j^{-1} ,
\nonumber
\\
   && \left(\begin{array}{c}
    B_\mu \\
    B_y \\
   \end{array} \right) (x,y_j-y)
   =
    \left(\begin{array}{c}
     B_\mu \\
     -B_y \\
   \end{array}\right) (x,y_j+y) ,
\nonumber
\\
 && P_j =\textrm{diag}(-1,-1,-1,-1,+1) ,
 \qquad (j=0,1),
 \label{pj}
\eea 
which reduce the $SO(5)\times U(1)_X$ symmetry to
$SO(4)\times U(1)_X$.
A scalar field $\Phi(x)$ on the Planck brane
belongs to $(0,{1\over 2})$ representation
of $SO(4)\sim SU(2)_L \times SU(2)_R$ and 
has a charge of $U(1)_X$.
With the brane action
\bea
  S_{\textrm{\scriptsize Pl. brane}}^{%
   \textrm{\scriptsize scalar}}
 &\!\!\!=\!\!\!&\int d^5x \delta(y)
  \left\{
     -(D_\mu \Phi)^\dag D^\mu \Phi
     -\lambda_\Phi
     (\Phi^\dag \Phi -w^2)^2 \right\} ,
\nonumber
\\
  D_\mu \Phi &\!\!\!=\!\!\!&
   \partial_\mu \Phi 
   -i\left(g_A \sum_{a_R}^3
     A_\mu^{a_R} T^{a_R}
      +{g_B\over 2}B_\mu \right) \Phi ,
      \label{cov}
\eea
the $SU(2)_R\times U(1)_X$ symmetry breaks
down to $U(1)_Y$,
the weak hypercharge in the standard model.
The massless modes
of $A_\mu^{1_R}$, $A_\mu^{2_R}$ and
$A_\mu^{'3_R}$ acquiring large masses.
Here
\bea
   \left(\begin{array}{c}
    A_M^{'3_R} \\
    A_M^Y \\
   \end{array}\right) 
   = \left(\begin{array}{cc}
    c_\phi & -s_\phi \\
    s_\phi & c_\phi \\
   \end{array}\right) 
  \left(\begin{array}{c}
    A_M^{3_R} \\
   B_M \\
  \end{array} \right) ,
 \quad
    c_\phi ={g_A\over \sqrt{g_A^2 + g_B^2}} ,
  \quad
    s_\phi ={g_B\over \sqrt{g_A^2 + g_B^2}} .
\eea
The four-dimensional gauge coupling for electromagnetic
interaction is
\bea
   e = {g_A g_B\over \sqrt{(g_A^2 +2g_B^2)L}} 
  ={g_A s_\phi\over \sqrt{(1+s_\phi^2)L}} .
\eea 
For $e=g\sin\theta_W$ and $g=g_A/\sqrt{L}$,
a relationship between the couplings and 
the weak mixing angle is
give by $s_\phi^2 =\tan^2\theta_W$.
We assume that $w$ is much larger than the 
KK mass scale, being of 
${\cal O}(M_{\textrm{\scriptsize GUT}})$
to ${\cal O}(M_{\textrm{\scriptsize Planck}})$.
The net effect for low-lying modes of the KK
towers of $A_\mu^{1_R}$, $A_\mu^{2_R}$
and $A_\mu^{'3_R}$ is that they effectively
obey Dirichlet boundary conditions at the Planck 
brane.
This is a limit of the Dirichlet-Neumann 
mixed boundary condition.
The effective orbifold boundary conditions are tabulated in Table~%
\ref{tab:bc}.
\begin{table}[htb]
\begin{center}
\caption{Boundary conditions for gauge bosons and bulk fermions:
The effective 
Dirichlet condition made by brane dynamics
is denoted as D${}_{\textrm{\scriptsize eff}}$.
For fermions $j=1,\cdots,4$.\label{tab:bc}} 
\vskip 7pt
\begin{tabular}{ccccccc}
\hline\hline
$A_\mu^{a_L}$ & $ A_\mu^{1_R,2_R}$ & $A_\mu^{'3_R}$ & $A_\mu^Y$ & $A_\mu^{\hat{a}}$ & $B_\mu$ 
& $\psi_{ajL},\psi_{a5R}$ \\ \hline
(N,N) & (D${}_{\textrm{\scriptsize eff}}$,N) & 
(D${}_{\textrm{\scriptsize eff}}$,N) & (N,N) 
& (D,D) & (N,N)  & (N,N) \\ \hline
$A_y^{a_L}$ & $ A_y^{1_R,2_R}$ & $A_y^{'3_R}$ & $A_y^Y$ & $A_y^{\hat{a}}$ & $B_y$ 
&$\psi_{ajR},\psi_{a5L}$ \\ \hline
(D,D) & (D,D) & (D,D) & (D,D) & (N,N) & (D,D)
& (D,D) \\ \hline
\end{tabular}
\end{center}
\end{table}
From the consistency requirement with 
the five-dimensional gauge transformation,
the effective Dirichlet condition in Table~%
\ref{tab:bc}
is not allowed to be replaced by Dirichlet condition 
without brane dynamics~\cite{Csaki:2003dt, Sakai:2006qi}.

Bulk fermions for quarks and leptons are introduced as
multiplets in 
the vectorial representation 
of $SO(5)$.
In the quark sector two vector multiplets are introduced for 
each generation. In the lepton sector 
it suffices to introduce one multiplet for each generation
to describe massless neutrinos, whereas
it is necessary to introduce two multiplets to
describe massive neutrinos.
They are denoted by 
$\Psi_a^t =(\psi_{a1},\ldots,\psi_{a5})^t$
where the subscript $a$ runs from 1 to 3 or 4
for each generation.

In the bulk the action integral is
\bea
   S_{\textrm{\scriptsize bulk}}^{%
     \textrm{\scriptsize fermion}}
   &\!\!\!=\!\!\!& \int d^5 x \sqrt{-G}
 \sum_{a=1}^3 i\bar{\Psi}_a {\cal D}(c_a) \Psi_a ,
\\
   {\cal D}(c_a)
    &\!\!\!=\!\!\!&
      \Gamma^A e_A^M
      (\partial_M +{1\over 8}\omega_{MBC}
       [\Gamma^B, \Gamma^C]
       -ig_A A_M
       -ig_B Q_{Xa} B_M)
     -c_a \sigma'(y) ,
     \label{sfermi}
\eea
where the Dirac conjugate is 
$\bar{\Psi} =i\Psi^\dag \Gamma^0$
and Gamma matrices are given by
\bea
  \Gamma^\mu = \left(\begin{array}{cc}
     & \sigma^\mu \\
     \bar{\sigma}^\mu & \\
     \end{array}\right) ,\quad
 \Gamma^5 =\left(\begin{array}{cc}
    1 & \\
     & -1 \\
     \end{array}\right) ,
   \quad
   \sigma^\mu= (1,\vec{\sigma}) , \quad
   \bar{\sigma}^\mu=(-1,\vec{\sigma}) .
\eea
The non-vanishing spin connection is
$\omega_{\mu m5} =-\sigma' e^{-\sigma}\delta_{\mu m}$, where 
$\delta_{\mu m}$ denotes a vierbein in
the four-dimensional Minkowski spacetime.
The $c_a$ term in Eq.~(\ref{sfermi}) gives
a bulk kink mass, where
$\sigma'(y)=k\epsilon(y)$ is a periodic
step function with a magnitude $k$.
The dimensionless parameter $c_a$ plays an important 
role in the Randall-Sundrum warped spacetime.
The orbifold boundary conditions are given by
\bea
  \Psi_a (x,y_j-y) &\!\!\!=\!\!\!& P_j \Gamma^5 \Psi_a (x,y_j+y) .
\eea
With $P_j$ in Eq.~(\ref{pj}) the first four component of
$\Psi_a$ are even under parity for the 4D left-handed 
$(\Gamma^5=-1)$ components.
An $SO(5)$ vector $\Psi$ can be expressed
as the sum of
$({1\over 2},{1\over 2})$ representation and a singlet $(0,0)$ of $SU(2)_L\times SU(2)_R$.
The $({1\over 2},{1\over 2})$ representation is written
as
\bea
   \hat{\psi}&\!\!\!=\!\!\!&
    \left(\begin{array}{cc}
      \hat{\psi}_{11} & \hat{\psi}_{12} \\
      \hat{\psi}_{21} & \hat{\psi}_{22} \\
      \end{array} \right)
  =
  {1\over \sqrt{2}}
   (\psi_4 +i\vec{\psi}\cdot\vec{\sigma})
   i\sigma_2
  =-{1\over \sqrt{2}}
  \left( \begin{array}{cc}
     \psi_2 +i\psi_1 & -(\psi_4 +i\psi_3) \\
     \psi_4 -i\psi_3 &  \psi_2 -i\psi_1 \\
  \end{array}\right) .
\eea 
The singlet (0,0) is $\psi_5$.
The quarks in the third generation, for instance,
are composed of bulk Dirac fermions of $SO(5)$ vectorial
representation
\bea
\Psi_1  \left({2\over 3}\right)
 &\!\!\!=\!\!\!&
 \left[Q_1 =\left(\begin{array}{c}
     T \\
     B \\
     \end{array}\right) ,~
    q=\left(\begin{array}{c}
     t \\
     b \\
     \end{array}\right) ,~
   t' \right] ,
\\
 \Psi_2 \left(-{1\over 3}\right)
 &\!\!\!=\!\!\!&
 \left[Q_2 =\left(\begin{array}{c}
     U \\
     D \\
     \end{array}\right) ,~
    Q_3=\left(\begin{array}{c}
     X \\
     Y \\
     \end{array}\right) ,~
   b' \right] ,  
\eea
and boundary right-handed fermions of
the $({1\over 2},0)$
representation for $SU(2)_L \times SU(2)_R$
\bea
   \hat{\chi}_{1R} 
 \left({7\over 6}\right)
  =
    \left(\begin{array}{c}
       \hat{T}_R \\
       \hat{B}_R \\
       \end{array}\right) ,  \qquad
   \hat{\chi}_{2R} 
  \left({1\over 6}\right)
   =
    \left(\begin{array}{c}
       \hat{U}_R \\
       \hat{D}_R \\
       \end{array}\right) , \qquad
\hat{\chi}_{3R}
  \left(-{5\over 6}\right)
   =
    \left(\begin{array}{c}
       \hat{X}_R \\
       \hat{Y}_R \\
       \end{array}\right) .    
\eea   
For boundary fermions,
the hypercharge $Y/2$ is equal to 
the $U(1)_X$ charge, $Q_X$.
The leptons in the third generation
are
composed of bulk Dirac fermions of $SO(5)$ vectorial
representation
\bea
   \Psi_3 
  (-1)
  &\!\!\!= \!\!\! &
   \left[
   \ell =\left(\begin{array}{c}
      \nu_\tau \\
      \tau \\
      \end{array}\right) ,~
    L_1 =\left(\begin{array}{c}
      L_{1X} \\
      L_{1Y} \\
      \end{array}\right) ,~
      \tau'\right] ,
      \label{psi3}
\\
   \Psi_4 (0) &\!\!\!=\!\!\!&
     \left[
        L_2 = \left(\begin{array}{c}
            L_{2X} \\
            L_{2Y} \\
            \end{array}\right) , ~
        L_3 =\left(\begin{array}{c}
             L_{3X} \\
             L_{3Y} \\
              \end{array}\right) , ~
        \nu'_{\tau} \right] ,
        \label{psi4}
\eea
and boundary right-handed fermions of
the $({1\over 2},0)$ representation for $SU(2)_L \times SU(2)_R$
\bea
  \hat{\chi}_{1R}^\ell
  \left(-{3\over 2}\right) =\left(\begin{array}{c}
     \hat{L}_{1XR} \\
     \hat{L}_{1YR} \\
     \end{array}\right) , 
\quad
     \hat{\chi}_{2R}^\ell
  \left({1\over 2}\right) =\left(\begin{array}{c}
     \hat{L}_{2XR} \\
     \hat{L}_{2YR} \\
     \end{array}\right) , 
\quad
     \hat{\chi}_{3R}^\ell
  \left(-{1\over 2}\right) =\left(\begin{array}{c}
     \hat{L}_{3XR} \\
     \hat{L}_{3YR} \\
     \end{array}\right) .   
\eea
The number in the parenthesis on the left-hand side
for each fermion
denotes the $U(1)_X$ charge.
The hypercharge and
the electric charge are 
given by
$Y/2=T^{3_R} +Q_X$ and 
$Q_E=T^{3_L}+T^{3_R}+Q_X$, respectively. 
The components $B$ and $t$ couple to $t'$ through
the vacuum expectation value $\langle A_y^c
\rangle \propto \theta_H T^{\hat{4}}$ with
the Wilson line phase $\theta_H$. 
Similarly, 
$D$ and $X$ couple to $b'$ and for leptons
$\tau$ and $L_{1X}$ couple to $\tau'$.
With $\theta_H$ alone, there remain extra massless
modes of fermions.
In order to make them heavy,
we introduce right-handed fermion $\hat{\chi}_{\alpha R}$
and $\hat{\chi}_{\alpha R}^\ell$
in the $({1\over 2},0)$ representations of 
$SU(2)_L\times SU(2)_R$
localized on the Planck brane $y=0$.
The brane fermions $\hat{\chi}_{\alpha R}$ and
$\hat{\chi}_{\alpha R}^\ell$ couple to the corresponding
bulk fermions and the brane scalar $\Phi$ in Eq.~(\ref{cov})
through Yukawa couplings.
After the $\Phi$ 
develops the vacuum expectation value,
the general brane action for $\hat{\chi}_{\alpha R}$
and $\hat{\chi}_{\alpha R}^\ell$ is given by
\bea
 &&  S_{\textrm{\scriptsize Pl.brane}}^{%
   \textrm{\scriptsize fermion}}
   = \int d^5 x \, \sqrt{-G} \,
   i\delta(y)
\nonumber
\\
  &&\quad \times
   \left\{
   \sum_{\alpha=1}^3 \left[
   \hat{\chi}_{\alpha R}^\dag
   \bar{\sigma}^\mu
    D_\mu \hat{\chi}_{\alpha R}
    - \mu_\alpha
    (\hat{\chi}_{\alpha R}^\dag Q_{\alpha L}
    -Q_{\alpha L}^\dag \hat{\chi}_{\alpha R})
      \right]
  -\tilde{\mu} (\hat{\chi}_{2R}^\dag q_L
  -q_L^\dag \hat{\chi}_{2R})
 \right. 
\nonumber
\\
  &&\quad \left.   
  + \sum_{\alpha=1}^3 \left[
   \hat{\chi}_{\alpha R}^{\ell \dag}
   \bar{\sigma}^\mu
    D_\mu \hat{\chi}_{\alpha R}^\ell
    - \mu_\alpha^\ell
    (\hat{\chi}_{\alpha R}^{\ell\dag} L_{\alpha L}
    -L_{\alpha L}^\dag \hat{\chi}_{\alpha R}^\ell)
      \right]
  -\tilde{\mu}^\ell (\hat{\chi}_{3R}^{\ell\dag} \ell_L
  -\ell_L^\dag \hat{\chi}_{3R}^\ell)
 \right\} ,
\eea
where $D_\mu$ in the kinetic term
has the same form as in Eq.~(\ref{cov}) 
with $A_\mu^{a_R}T^{a_R}$ replaced by
$A_\mu^{a_L}T^{a_L}$.
The $\mu$ terms mix bulk left-handed fermions and brane 
right-handed fermions.
When each coupling $\mu_{\alpha}$, $\tilde{\mu}$,
$\mu_\alpha^\ell$ and $\tilde{\mu}^\ell$ is
taken as a matrix, a flavor mixing can be introduced~\cite{%
Burdman:2002se, Uekusa:2008iz}.
These couplings have the dimension $[M]^{1/2}$
and are collectively denoted as $\mu$.
For simplicity, we adopt flavor-diagonal couplings.
Then only the modest conditions,
$\mu^2 \gg m_{\textrm{\scriptsize KK}}
 \equiv \pi k/(z_L-1)$,
 are necessary to  be satisfied
for the values of the couplings to get the desired low
energy mass spectrum.
The brane scalar $\Phi$ is not required to 
be protected from having the quadratic divergent
corrections.
In the case neutrinos are massless, $\Psi_4$, 
$\hat{\chi}_{2R}^\ell$ and $\hat{\chi}_{3R}^\ell$
are unnecessary. 
The model is anomaly free
with respect to
$SU(2)_L\times SU(2)_R \times U(1)_X$
independently of inclusion of these three fields.

\section{Z boson couplings of quarks and leptons \label{zcurrent}}

In order to evaluate the forward-backward 
production asymmetry on the $Z$ resonance for quarks
and leptons, the 
four-dimensional gauge couplings of fermions with $Z$ boson
are needed.
This derivation is given in Ref.~\cite{HNU}.
In this section, we summarize the resulting $Z$ boson couplings.

The four-dimensional Lagrangian terms 
for the $Z$ boson coupling of $t$
quark are obtained as
\bea
 {\cal L}_{Zt\bar{t}} = {g_A Z_\mu\over \cos\theta_W} 
  \left({\cal T}_{L}
  \bar{t}_L \gamma^\mu t_L
 +{\cal T}_{R}
   \bar{t}_R \gamma^\mu t_R \right) ,
    \label{zt}
\eea 
with
${\cal T}_{L,R}=
{1 \over 2}{\cal T}^{3}_{L,R}
-{2\over 3} {\cal T}^{Q}_{L,R} \sin^2 \theta_W$.
The quantities ${\cal T}_L^3$ and ${\cal T}_L^Q$ are
given by
\bea
  {\cal T}_{L}^3
   &\!\!\!=\!\!\!& \int_1^{z_L} dz \,
      \bigg[ N_Z(z) (C_L(z;\lambda_t))^2\left(
       a_U^2
           -2\cos\theta_H a_{B+t} a_{B-t} 
   \right)
\nonumber
\\
   && 
     -2\sin\theta_H D_Z(z) C_L(z;\lambda_t) S_L(z;\lambda_t)
     a_{B+t} a_{t'}
 \bigg] ,
   \label{tl3}
\\
  {\cal T}_{L}^Q
 &\!\!\!=\!\!\!&
    \int_1^{z_L} dz \,  N_Z 
  \bigg[(C_L(z;\lambda_t))^2
   (a_U^2 +a_{B+t}^2 +a_{B-t}^2)
   +(S_L(z;\lambda_t))^2
     a_{t'}^2
    \bigg] .
    \label{tlq}
\eea
Here $N(z)$ and $D(z)$ are the fundamental
functions in mode expansion of $Z$ boson,
$C_L(z;\lambda_t)$ and $S_L(z;\lambda_t)$ are
the fundamental functions in mode expansion of $t$ quark.
In expressing mode function profiles,
the conformal coordinate $z=e^{\sigma(y)}$
for the fifth dimension is employed, with which
the metric becomes
$ds^2 =z^{-2} \{\eta_{\mu\nu} dx^\mu dx^\nu + dz^2/k^2\}$.
The fundamental region $0\leq y\leq L$ is 
mapped to $1\leq z \leq z_L=e^{kL}$.
In the bulk region $0<y<L$, one has  
$\partial_y =kz \partial_z$,
$A_y =kz A_z$, $B_y = kz B_z$.
The $a_U$, $a_{B+t}$, $a_{B-t}$ and $a_{t'}$ are
the coefficients in mode expansion of $t$ quark.
These explicit definitions are shown in Appendix~\ref{app:mode}.
For the right-handed $t$ quark,
${\cal T}_R^3$ and ${\cal T}_R^Q$ are
given by ${\cal T}_L^3$ and ${\cal T}_L^Q$
with $(C_L,S_L)$ replaced by $(S_R,C_R)$, respectively.

The four-dimensional Lagrangian terms for the $Z$ boson 
coupling of $b$ quark are
\bea
 {\cal L}_{Zb\bar{b}}
  = {g_A Z_\mu\over \cos\theta_W} 
  \left({\cal B}_{L}
  \bar{b}_L \gamma^\mu b_L
 +{\cal B}_{R}
   \bar{b}_R \gamma^\mu b_R \right) ,
\eea 
with
${\cal B}_{L,R}=
-{1 \over 2}{\cal B}^{3}_{L,R}
+{1\over 3} {\cal B}^{Q}_{L,R}\sin^2 \theta_W$,
where ${\cal B}_L^3$ and ${\cal B}_L^Q$ are
given by
\bea
  {\cal B}_{L}^3
   &\!\!\!=\!\!\!& \int_1^{z_L} dz \,
      \bigg[ N_Z(z) (C_L(z;\lambda_b))^2\left(
       a_b^2
           +2\cos\theta_H a_{D+X} a_{D-X} 
   \right)
\nonumber
\\
   && 
     +2\sin\theta_H D_Z(z) C_L(z;\lambda_b) S_L(z;\lambda_b)
     a_{D+X} a_{b'}
 \bigg] ,
\\
  {\cal B}_{L}^Q
 &\!\!\!=\!\!\!&
     \int_1^{z_L} dz \,  N_Z(z) 
  \bigg[(C_L(z;\lambda_b))^2
   (a_b^2 +a_{D+X}^2 +a_{D-X}^2)
   +(S_L(z;\lambda_b))^2
     a_{b'}^2
    \bigg] .
\eea
For the right-handed $b$ quark,
${\cal B}_R^3$ and ${\cal B}_R^Q$ are
given by ${\cal B}_L^3$ and ${\cal B}_L^Q$
with $(C_L,S_L)$ replaced by $(S_R,C_R)$, respectively.

To describe a massless neutrino for each generation,
we need to introduce only a vector multiplet $\Psi_3$ in 
Eqs.~(\ref{psi3}) and (\ref{psi4}).
The four-dimensional 
Lagrangian terms for the $Z$ boson coupling of $\nu_\tau$
neutrino are obtained as
\bea
 {\cal L}_{Z\nu\bar{\nu}}
  = {g_A Z_\mu\over \cos\theta_W} 
 {\cal N}_{L}
  \bar{\nu}_{\tau L} \gamma^\mu \nu_{\tau L}
  ,
    \label{znu}
\eea 
with
${\cal N}_{L}=
{1 \over 2}{\cal N}^{3}_{L}$,
where ${\cal N}_L^3$ is
given by
\bea
  {\cal N}_{L}^3
   &\!\!\!=\!\!\!& \int_1^{z_L} dz \,
      \bigg[ N_Z(z) (C_L(z;0))^2
       a_{\nu_\tau}^2
 \bigg] .
   \label{nl3}
\eea
The four-dimensional 
Lagrangian terms for the $Z$ boson coupling of $\tau$ lepton 
are
\bea
  {\cal L}_{Z\tau\bar{\tau}}
 ={g_A Z_\mu\over \cos\theta_W} 
  \left({\cal X}_{L}
  \bar{\tau}_L \gamma^\mu \tau_L
 +{\cal X}_{R}
   \bar{\tau}_R \gamma^\mu \tau_R \right) ,
\eea 
with
${\cal X}_{L,R}=
-{1 \over 2}{\cal X}^{3}_{L,R}
+{\cal X}^{Q}_{L,R}\sin^2 \theta_W$.
The quantities ${\cal X}_L^3$ and ${\cal X}_L^Q$ are
given by
\bea
  {\cal X}_{L}^3
   &\!\!\!=\!\!\!& \int_1^{z_L} dz \,
      \bigg[ N_Z(z) (C_L(z;\lambda_\tau))^2\left(
        2\cos\theta_H a_{\tau+L_{1X}} a_{\tau-L_{1X}} 
   \right)
\nonumber
\\
   && 
     +2\sin\theta_H D_Z(z) C_L(z;\lambda_\tau) 
  S_L(z;\lambda_\tau)
     a_{\tau+L_{1X}} a_{\tau'}
 \bigg] ,
\\
  {\cal X}_{L}^Q
 &\!\!\!=\!\!\!&
     \int_1^{z_L} dz \,  N_Z(z) 
  \bigg[(C_L(z;\lambda_\tau))^2
   (a_{\tau+L_{1X}}^2 +a_{\tau-L_{1X}}^2)
   +(S_L(z;\lambda_\tau))^2
     a_{\tau'}^2
    \bigg] .
\eea
For the right-handed $\tau$ lepton,
${\cal X}_R^3$ and ${\cal X}_R^Q$ are
given by ${\cal X}_L^3$ and ${\cal X}_L^Q$
with $(C_L,S_L)$ replaced by $(S_R,C_R)$, respectively.

To describe massive neutrinos one needs to introduce two multiplets
$\Psi_3$ and $\Psi_4$.  The structure is the same as 
in the quark sector.
The quantities ${\cal N}_{L,R}^3$ and
${\cal X}_{L,R}^{3,Q}$
are obtained with use of
the correspondence between leptons and quarks:
\bea
 (\nu_\tau,L_{2Y}, L_{3X}, \nu'_\tau)
  &\!\!\!\leftrightarrow\!\!\!&
(U, B, t, t'),
  \qquad
(\hat{L}_{3XR}, \hat{L}_{2YR})
  \leftrightarrow
(\hat{U}_R, \hat{B}_R),
\nonumber
\\
(L_{3YL},\tau,L_{1X}, \tau')
 &\!\!\!\leftrightarrow\!\!\!&
(b, D, X, b') ,
\qquad
(\hat{L}_{3YR}, \hat{L}_{1XR})
 \leftrightarrow
(\hat{D}_R, \hat{X}_R) ,
\nonumber
\\
(\mu_1^\ell, \mu_2^\ell, \mu_3^\ell, \tilde{\mu}^\ell)
 &\!\!\!\leftrightarrow\!\!\!&
(\mu_3, \mu_1, \tilde{\mu}, \mu_2) .
\eea
The four-dimensional Lagrangian term for the $Z$ boson coupling of 
the right-handed $\nu_\tau$ neutrino should be added as
\bea
  {\cal L}_{Z\nu\bar{\nu} R}
  =  {g_A Z_\mu \over \cos\theta_W} {\cal N}_R 
      \bar{\nu}_{\tau R} \gamma^\mu \nu_{\tau R} ,
\eea
with ${\cal N}_R ={1\over 2}{\cal N}_R^3$.

Typical numerical values for these $Z$ couplings 
have been given in Ref.~\cite{HNU}.
It has been shown that they deviate slightly from 
the standard model.
In next section, we will calculate $A_{FB}$
as an observed indication for the $Z$ couplings.

\section{Forward-backward asymmetry: numerical analysis \label{afb}}

A formal equation for 
the forward-backward asymmetry on the $Z$ resonance
for the scattering
$e^+ e^- \to f\bar{f}$ is the same
as in the standard model at tree level.
For the unpolarized initial electron, 
it is given by a simple formula
\bea
  A_{FB}^f ={3\over 4} \,
   A_{LR}^e A_{LR}^f .
    \label{afbform}
\eea   
Here the superscript 
denotes spices of fermions such as $e$ as in $A_{LR}^e$.
The polarization asymmetry for the decay
$Z\to f\bar{f}$ is determined from the $Z$ boson couplings
of fermions.
The $A_{LR}^f$ is given by
\bea
  A_{LR}^f =
    {(g_L^f)^2 -(g_R^f)^2 \over (g_L^f)^2 +(g_R^f)^2}
  .
   \label{eq:alr}
\eea
Here
$g_{L,R}^f$ are
${\cal T}_{L,R}$, ${\cal B}_{L,R}$, ${\cal N}_{L,R}$,
${\cal X}_{L,R}$ for $t$ and $b$ quarks
and $\nu_\tau$ and $\tau$ leptons, respectively.
In the standard model, at tree level $g_{L,R}^f$ are read as
\bea
   {\cal T}_L
 &\!\!\! \stackrel{\textrm{\scriptsize SM}}{\to}
  \!\!\!& \textrm{${1\over 2}$}
  -\textrm{${2\over 3}$} \sin^2 \theta_W ,
  \quad
   {\cal B}_L
   \stackrel{\textrm{\scriptsize SM}}{\to}
   -\textrm{${1\over 2}$} +
  \textrm{${1\over 3}$} \sin^2\theta_W ,
 \quad
  {\cal N}_L
  \stackrel{\textrm{\scriptsize SM}}{\to}
   \textrm{${1\over 2}$} ,
  \quad
  {\cal X}_L 
  \stackrel{\textrm{\scriptsize SM}}{\to}
  -\textrm{${1\over 2}$} +\sin^2\theta_W ,
\nonumber
\\
  {\cal T}_R 
  &\!\!\! \stackrel{\textrm{\scriptsize SM}}{\to}
  \!\!\!&
  -\textrm{${2\over 3}$}\sin^2\theta_W ,
 \qquad
  {\cal B}_R 
  \stackrel{\textrm{\scriptsize SM}}{\to} 
   \textrm{${1\over 3}$}\sin^2\theta_W ,
  \qquad\quad~
  {\cal N}_R
  \stackrel{\textrm{\scriptsize SM}}{\to}
   0 ,
 \quad
   {\cal X}_R 
    \stackrel{\textrm{\scriptsize SM}}{\to} \sin^2 \theta_W ,
\label{smt}
\eea
which are independent of the generation for fermions. 
In the present model, $g_{L,R}^f$ are defined
as quantities including integrals with respect to $z$
as given in the previous section.
The asymmetries $A_{LR}^f$ and $A_{FB}^f$ are obtained
through a numerical calculation.

The model with massive neutrinos has 16 parameters:
$c$ (whose number is 6) for three generations of quarks and leptons,
$|\tilde{\mu}/\mu_2|$ (whose number is 3)
for three generations of quarks,
$|\mu_3^\ell/\tilde{\mu}^\ell|$ (whose number is 3)
for three generations of leptons, and
$g_A$, $g_B$, $k$, $L$ except for
the modestly-tuned parameters,
$\mu^2 \gg m_{\textrm{\scriptsize KK}}$,
to make unwanted fields heavy.
We take 16 inputs as
6 quark masses, 6 lepton masses,
$\sin\theta_W$, $m_Z$, $k$ and $z_L$.
If one treats neutrinos as massless approximately,
the number of the parameters is reduced. 
The model with massless neutrinos has 13 parameters:
$c$ (whose number is 6) for three generations of quarks and leptons,
$|\tilde{\mu}/\mu_2|$ (whose number is 3)
for three generations of quarks,
and
$g_A$, $g_B$, $k$, $z_L$.
For this case, the 13 inputs can be taken as
6 quark masses, 3 charged lepton masses,
$\sin\theta_W$, $m_Z$, $k$ and $z_L$.
The case with massless neutrinos can also be regarded as a limit of 
the case with massive neutrinos.
The parameters $k$ and $z_L$ are inputs for warped geometry,
$\sin\theta_W$ and $m_Z$ are inputs for the $Z$ boson
and the others are assigned for masses of fermions.

For input parameters of warped geometry, we take
$z_L= (10^{18}\textrm{GeV})/(1\textrm{TeV}) 
=1.0 \times 10^{15}$ and
$k=4.7 \times 10^{17}$, where $k$ is chosen
such that the value of $m_W=m_W(k,z_L,\theta_H)$ is
appropriately reproduced.
For these values of $z_L$ and $k$, 
the Kaluza-Klein scale is given by
$m_{\textrm{\scriptsize KK}}=\pi k z_L^{-1} =1.48$TeV.
For input parameters of the $Z$ boson,
we take 
$m_Z =91.1876$GeV (the central values in Particle 
Data Group data~\cite{PDG}) and
$\sin^2 \theta_W =0.2312$
(the $\overline{\textrm{MS}}$ value in
Particle Data Group data~\cite{PDG}).
The Wilson line phase is
$\theta_H=\pi/2$.
\begin{table}[htb]
\begin{center}
\caption{
Masses of quarks and charged leptons:
the central values in unit of MeV in 
XZZ~\cite{XZZ}, FK~\cite{FK}, PDG~\cite{PDG}.
\label{tab:qpdg}
}
\vskip 5pt
\begin{tabular}{cccccccccc}
\hline \hline &
$m_u$ & $m_d$ &
$m_s$ & $m_c$ &
$m_b$ & $m_t$ &
$m_e$ & 
$m_\mu$ &
$m_\tau$ \\
\hline XZZ &
1.27 & 2.90 &
55 & 619 &
2890 & 171700 &
0.486570161 & 
102.7181359 &
1746.24 \\  
FK &
2.33 & 4.69 &
93.4 & 677 &
3000 & 181000 &
0.48684727 & 
102.75138 &
1746.69 \\ 
PDG &
2.4 & 4.75 &
104 & 1270 &
4200 & 171200 &
0.510998910 & 
105.658367 &
1776.84 \\
\hline
\end{tabular}
\end{center}
\end{table}
For fermion masses, we take three sets: 
The first set is
the central values of the running masses 
at the $m_Z$ scale given in Ref.~\cite{XZZ}, which 
we refer to as XZZ.
The second set is 
the central values of the running masses
at the $m_Z$ scale given in Ref.~\cite{FK}, which
we refer to as FK.
The final set is 
the central values of the physical (pole) masses
given in Ref.~\cite{PDG}, which
we refer to as PDG.
The mass parameters of the three sets
are tabulated in Table~\ref{tab:qpdg}.
In the sets XZZ and FK, neutrinos are taken as massless.
In the set PDG, 
$m_{\nu_e}=10^{-3}$eV, and
$m_{\nu_e}=m_{\nu_1}$, 
$m_{\nu_\mu} =m_{\nu_2}$,
$m_{\nu_\tau} =m_{\nu_3}$ are assumed. 
The differences of masses squared are taken as 
$\Delta m_{21}^2 =8 \times 10^{-5} \textrm{eV}^2$ and 
$\Delta m_{32}^2 =2.45 \times 10^{-3} \textrm{eV}^2$ 
(the central values for the difference of masses squared 
in Ref.~\cite{PDG}).
This corresponds to taking 
$m_{\nu_\mu}=9 \times 10^{-3}$eV and
$m_{\nu_\tau} =5.0309 \times 10^{-2}$eV.

With the above input parameters, 
the polarization asymmetries for the decay
$Z\to f\bar{f}$, $A_{LR}^f$
are numerically calculated. These values are
tabulated in Table~\ref{tab:alr}.
For comparison, the values derived from Eq.~(\ref{smt}) are shown
as the tree-level values
in the standard model. 
\begin{table}[htb]
\begin{center}
\caption{$A_{LR}^f$. \label{tab:alr}}
\vskip 3pt
\begin{tabular}{c|ccc|ccc|ccc}
\hline\hline &
$u$& $c$& $t$& $d$& $s$& $b$& $e$& $\mu$& $\tau$ \\
\hline
XZZ & 
0.6643 & 
0.6645 & 
0.5097 & 
0.9448 & 
0.9348 & 
0.9350 & 
0.1419 & 
0.1422 & 
0.1423   
\\
FK &
0.6643 & 
0.6645 & 
0.4890 & 
0.9348 & 
0.9348 & 
0.9350 & 
0.1419 & 
0.1422 & 
0.1423   
\\
PDG &
0.6643 & 
0.6645 & 
0.5108 & 
0.9348 & 
0.9348 & 
0.9350 & 
0.1418 & 
0.1420 & 
0.1421   
\\ \hline
SM &
0.6686 &  
&
&
0.9357 & 
&
&
0.1496 & 
&
\\
\hline
\end{tabular}
\end{center}
\end{table}
For neutrinos, $A_{LR}^\nu=1$, where the difference $A_{LR}^\nu-1$
is suppressed by large orders of magnitude.
It is read from Table~\ref{tab:alr} that
the values of $A_{LR}^f$ are close to the values of
the standard model.
On the other hand,
the flavor universality is slightly violated
except for $t$ quark.
The non-universality between $A_{LR}^u$ and $A_{LR}^t$
is estimated as $A_{LR}^t/A_{LR}^u -1 \sim -23\%$
for the set XZZ.

From the above values for $A_{LR}^f$ and the formula
(\ref{afbform}), 
the forward-backward asymmetries 
on the $Z$ resonance for $e^+e^- \to 
f\bar{f}$, $A_{FB}^f$ are calculated.
The values of them are tabulated in
Table~\ref{tab:afb}.
\begin{table}[htb]
\begin{center}
\caption{$A_{FB}^f$. For comparison, 
the values derived with use of Eq.~(\ref{smt}) are shown
as the tree-level values
in the standard model.
In addition, 
the central values and the deviations
of the standard model prediction given in 
Particle Data Group review~\cite{PDG} are shown
in parentheses. For XZZ, the deviation between
the prediction and the experimental value is denoted as Pull
$\equiv$ [(Central value)${}_{\textrm{\scriptsize Exp.}}-$(Prediction)]/(Error)${}_{%
\textrm{\scriptsize Exp.}}$.
Since $m_t>m_Z$, $A_{FB}^t$ should be interpreted
as a mere reference value given in 
Eq.~(\ref{afbform}) with $A_{LR}^f$.
\label{tab:afb}}
\vskip 7pt
\begin{tabular}{cccccc}
\hline\hline
$f$ & Exp. & XZZ& Pull & FK & PDG \\
\hline
$u$ &
--- &
0.07071 & 
&
0.07071 & 
0.07063   
\\
$c$ &
$0.0707\pm 0.0035$ &   
0.07073 & 
0.0 &
0.07073 & 
0.07065   
\\
$t$ &
--- &
0.05425 & 
&
0.05205 & 
0.05431   
\\
\hline
$d$ &
--- &
0.09950 & 
&
0.09950 & 
0.09939   
\\
$s$ &
$0.0976\pm 0.0114$ & 
0.09950 & 
-0.2 & 
0.09950 & 
0.09939   
\\
$b$ &
$0.0992\pm 0.0016$ & 
0.09952 & 
-0.2 & 
0.09952 & 
0.09941   
\\ \hline
$e$ &
$0.0145\pm 0.0025$ & 
0.01511 & 
-0.2  & 
0.01511 & 
0.01507   
\\
$\mu$ &
$0.0169\pm 0.0013$ & 
0.01513 & 
1.4  & 
0.01513 & 
0.01510   
\\     
$\tau$ &
$0.0188\pm 0.0017$ & 
0.01515 & 
2.1  & 
0.01515 & 
0.01511   
\\
\hline
\end{tabular}
\begin{tabular}{cc }
\hline\hline
SM & Pull \\ 
\hline
0.07500 & 
\\
(0.0738) &
(-0.9) \\  
&\\           
\hline
0.10496& \\  
(0.1034)&
(-0.5)
 \\   
(0.1033) &
(-2.5)\\   
\hline
0.01677 (0.01627)&
(-0.7) \\ 
& (0.5)\\           
& (1.5)\\           
\hline
\end{tabular}
\end{center}
\end{table}
It is found that the tree-level predictions of 
$A_{FB}^c$, $A_{FB}^s$ and $A_{FB}^b$
are quite close to the central values of the 
experimental values given in Ref.~\cite{PDG}.
This is a remarkable result of the present model,
which is different from the standard model.
As for the lepton sector,
it seems that both of theory and 
experiment need improvements.
The notable agreement in the quark sector is the case for
all the three sets, XZZ, FK, PDG of fermion masses.
That the $A_{FB}^f$ are not very sensitive to whether the 
fermion masses
are running masses or 
physical masses is led to relaxing the dependence 
of the predictions on
the values of $c$, $\mu_\alpha$, $\tilde{\mu}$.
Hence, the model have made a realistic 
prediction with a moderate tuning as a whole.

The polarization asymmetry and the forward-backward 
asymmetry are not very sensitive to the values of 
input parameters for warped geometry.
For a large warp factor $z_L =1.0 \times 10^{17}$
and $k=5.0\times 10^{19}$GeV,
$A_{LR}^f$ and $A_{FB}^f$ are
given in Appendix~\ref{app:wi}.
For these input parameters,
the predictions are also shown to be close to the central values of 
the experimental data.

\vspace{2ex}

\section{Other electroweak quantities
\label{sec:other}}

In addition to the forward-backward asymmetry,
the experimental measurement 
has been developed for other 
electroweak quantities.
In this section, we present
the tree-level prediction
of the decay width of $Z$ boson and
the $S$ and $T$ parameters.
For various electroweak quantities,
it has been proposed that there
is a class of theory beyond 
the standard model which is applied to
a global analysis~\cite{%
Barbieri:2004qk}.
Such an analysis may be useful when
the treatment of the model with
radiative corrections is as transparent
as in the standard model.
We will choose the values of
the input parameters given in the 
previous section instead of
searching a global fit.
One reason is because
the warp factor and the masses are fixed from the
hierarchy and the experiments, respectively
at the leading level.
The other is because the parameter values for
a global fit
would be affected by 
radiative corrections.
Our standpoint is that
the tree-level analysis should be prioritized
and that the quantitative results should be given.

\subsection{Decay width}

At tree level in the $SO(5)\times U(1)_X$ model,
the decay width of $Z$ boson is given by
\bea
   \Gamma( Z \to f\bar{f} )
   ={m_Z \alpha L \over 3 
    \sin^2 \theta_W \cos^2 \theta_W }
  \left[ {(g_L^f)^2
      + (g_R^f)^2 \over 2}
      + 2 g_L^f g_R^f
        {m_f^2 \over m_{Z}^2} \right]
      \sqrt{1-{4 m_f^2 \over m_Z^2}} . 
\eea
where the couplings $g_L^f$ and $g_R^f$ have been used
as in Eq.~(\ref{eq:alr}).
\begin{table}[htb]
\begin{center}
\caption{$Z$ boson decay:
the branching fraction and the
total width. For the values of masses, 
the set XZZ is adopted and $\alpha =1/128$. 
The invisible decay mode means the decay into 
$\nu_e \bar{\nu}_e + 
\nu_\mu \bar{\nu}_\mu + 
\nu_\tau \bar{\nu}_\tau$ in the model.
\label{tab:decay}}
\begin{tabular}{c|ccc}
\hline\hline
$Z$ decay modes & Fraction ($\Gamma_i/\Gamma$) & Exp.
& Pull
\\ \hline
$e^+ e^-~(\%)$ & 3.46173
& $3.363 \pm 0.004$ 
& -25
\\
$\mu^+ \mu^-~(\%)$ & 3.46097
& $3.366 \pm 0.007$
& -14
\\
$\tau^+\tau^-~(\%)$ & 3.45545
& $3.370 \pm 0.008$
& -11
\\ \hline
invisible $(\%)$ & 20.5164
& $20.00 \pm 0.06$
& -8.6
\\ \hline
$(u\bar{u} + c\bar{c})/2~(\%)$ & 11.7955
& $11.6 \pm 0.6$
& -0.3
\\ 
$(d\bar{d} + s\bar{s} + b\bar{b})/3 ~(\%)$ & 15.1715
& $15.6 \pm 0.4$
& 1.1
\\
$c\bar{c} ~(\%)$ & 11.794
& $12.03 \pm 0.21$
& 1.1
\\
$b\bar{b} ~(\%)$ & 15.1661
& $15.21 \pm 0.05$
& 0.9
\\ \hline
total width (GeV)
& 2.47617
& $2.4952 \pm 0.0023$
& 8.3
\\ \hline 
\end{tabular}
\end{center}
\end{table}
The total width and the branching fraction
are shown in Table~\ref{tab:decay}.
Table~\ref{tab:decay} includes the experimental values
given in Particle Data Group data~\cite{PDG}.
It is seen that the deviation of the tree level
prediction from the experimental value
is significantly large for leptonic decay modes
and that fractions in the quark sector are
not so different from the experimental data.
This deviation in the lepton sector seems too large.
We have find that
a comparatively large deviation in the lepton sector
arises also in the forward-backward asymmetry
in the previous section
and that $A_{FB}^b$, $A_{FB}^c$ and $A_{FB}^s$ in
the quark sector are quite close to 
the central values of the experimental values.
On the other hand, the analysis here has been
to estimate
only the leading contribution.
Since the lepton sector has the large flavor
mixing, the mixing effect
might change the fraction in the lepton sector
even at the tree level.
This analysis will be left to future work.

\subsection{$S$ and $T$ parameters}

In electroweak physics, it is conventional to
represent 
the effect of new physics for observables by
the $S$ and $T$ parameters~\cite{%
Peskin:1990zt, 
Peskin:1991sw}
(see \cite{Csaki:2005vy} for review).
This is estimated for new physics after 
loop corrections in the standard model are taken 
into account.
The general form of the Lagrangian 
associated with oblique
corrections is 
${\cal L}_{\textrm{\scriptsize eff}}
= {\cal L}_{\textrm{\scriptsize SM}}
+\tilde{\cal L}_{\textrm{\scriptsize new}}$,
with ${\cal L}_{\textrm{\scriptsize SM}}
= {\cal L}_{\textrm{\scriptsize SM}}(\tilde{e},
\sin \tilde{\theta}_W, \tilde{m}_Z, \tilde{m}_W)$
and
\bea
   \tilde{\cal L}_{\textrm{\scriptsize new}}
   &\!\!\!  =\!\!\!&
  {\Pi_{\gamma\gamma}'(0) \over 4} 
   \tilde{F}_{\mu\mu}\tilde{F}^{\mu\nu}
 +{\Pi_{WW}'(0)\over 2} 
  \tilde{W}^+_{\mu\nu} \tilde{W}^{-\mu\nu}
  +{\Pi_{ZZ}'(0)\over 4} 
   \tilde{Z}_{\mu\nu}\tilde{Z}^{\mu\nu}
  + {\Pi_{\gamma Z}'(0)\over 2} 
  \tilde{F}_{\mu\nu}\tilde{Z}^{\mu\nu}
\nonumber
\\
  && - \Pi_{WW}(0) 
    \tilde{W}_\mu^+ \tilde{W}^{-\mu}
      -{\Pi_{ZZ}(0)\over 2}
   \tilde{Z}_{\mu}
         \tilde{Z}^\mu .
\eea   
where $\tilde{F}_{\mu\nu}$, $\tilde{Z}_{\mu\nu}$ and
$\tilde{W}_{\mu\nu}^{\pm}$ are the field strengths
for photon, $Z$ boson and $W$ boson, respectively
and the quantities
$\Pi'_{\gamma\gamma}(0)$, $\Pi'_{WW}(0)$,
$\Pi'_{ZZ}(0)$, $\Pi_{WW}(0)$ and $\Pi_{ZZ}(0)$
are assumed to be small.
For the Lagrangian 
${\cal L}_{\textrm{\scriptsize eff}}$,
the $S$ and $T$ parameters are given by
\bea
  \alpha S =
      4 \tilde{s}^2 \tilde{c}^2 
        \left(\Pi_{ZZ}'(0)- \Pi_{\gamma\gamma}'(0) 
   -{\tilde{c}^2 -\tilde{s}^2 \over 
  \tilde{c} \tilde{s}} \Pi_{\gamma Z}'(0) \right)
        ,
\qquad
  \alpha T = {\Pi_{WW}(0) \over \tilde{m}_W^2}
      - {\Pi_{ZZ}(0) \over \tilde{m}_Z^2} ,
\eea
where $\tilde{s}=\sin\tilde{\theta}_W$
and $\tilde{c}=\cos\tilde{\theta}_W$.
After the kinetic terms are canonically normalized,
the gauge boson part of
the effective Lagrangian becomes
\bea
    {\cal L}_{\textrm{\scriptsize eff}}
      &\!\!\!=\!\!\!&
        -{1\over 4} F_{\mu\nu} F^{\mu\nu}
          -{1\over 2} W_{\mu\nu}^+
            W^{-\mu\nu}
            -{1\over 4} Z_{\mu\nu} Z^{\mu\nu}         
  - m_W^2 W_\mu^+ W^{-\mu}
     -{1\over 2}  
    m_Z^2
       Z_\mu Z^\mu
\nonumber
\\
  &&
 + e \sum_i \bar{f}_i \gamma^\mu Q_E
              f_{i} A_\mu
   +\left[ {\tilde{e}\over \sqrt{2}\tilde{s}}
       \left(1+{\Pi_{WW}'(0)\over 2}\right)
         \sum_{ij} \tilde{V}_{ij}
           \bar{f}_i 
             \gamma^\mu P_L f_j W_\mu^+
               + \textrm{h.c.} \right]
\nonumber
\\
  &&
    +{\tilde{e}\over \tilde{s} \tilde{c}}
       \left(1+{\Pi_{ZZ}(0)\over 2}\right)
         \sum_i \bar{f}_i \gamma^\mu
           \left[
            T^3_i P_L
              -Q_E \tilde{s}^2
                + Q_E \tilde{s} \tilde{c}
               \Pi_{\gamma Z}'(0)\right] f_i Z_\mu .
       \label{smpnew}          
\eea
where $T_i^3$ is the eigenvalue of $T_L^3$ for 
a fermion $f_i$.
The physical masses and coupling are
identified as
$m_W = (1 + (\Pi_{WW}(0)/\tilde{m}_W^2
 +\Pi_{WW}'(0))/2) \tilde{m}_W$,
$m_Z = (1 + (\Pi_{ZZ}(0)/\tilde{m}_Z^2
 +\Pi_{ZZ}'(0))/2) \tilde{m}_Z$
and $e= (1+\Pi_{\gamma\gamma}'(0)/2) \tilde{e}$.
In Eq.~(\ref{smpnew}),
the effect of new physics is included in
the charged and neutral currents.
The $Z$ boson coupling in Eq.~(\ref{smpnew})
is written in terms of $S$ and $T$ as
\bea
        {e (1+\alpha T/2)
      \over s c}            
       \sum_i \bar{f}_i
       \gamma^\mu
       \left[ T^3_i P_L
       -Q_E \left(s^2
        +{\alpha S\over 4(c^2 -s^2)}
     -{c^2 s^2 \alpha T
      \over c^2 -s^2} \right) \right]
       f_i Z_\mu ,
         \label{zst}
\eea
where $s$ satisfies 
$e^2/(s^2 c^2 m_Z^2) =\tilde{e}^2 (1-\Pi_{WW}(0)/\tilde{m}_W^2)/(\tilde{s}^2 \tilde{c}^2
\tilde{m}_Z^2)$.
If an effective Lagrangian is in the form of
Eq.~(\ref{smpnew}), the corresponding
$S$ and $T$ can be estimated.
The $SO(5)\times U(1)_X$ model has 
gauge couplings of fermions
different from the values in the standard model
while the canonical kinetic terms are kept.
We have found that this occurs at tree level
and that this is the effect of
extra-dimensional new physics.
The point is 
that in the $SO(5)\times U(1)_X$
model the gauge interactions are not universal
with respect to the generation of fermions
unlike Eq.~(\ref{smpnew}).
Due to this deviation,
the tree-level currents in the $SO(5)\times
U(1)_X$ model may need some alternative
parameters instead of the $S$ and $T$ 
parameters.
However,
we will not discuss this issue further.
Our tree-level estimation is to find how large
the values are
if contributions of the $S$ and $T$ parameters rather
than the alternative parameters
were dominant for corrections to couplings.
In other words, for the moment,
we treat the case where corrections are 
characterized by the $S$ and $T$ parameters
with a flavor fixed.

We formally estimate the tree-level values of 
the $S$ and $T$ parameters in the $SO(5)\times U(1)_X$
model.
For a fermion $f$ with corrections
dominated by the $S$ and $T$ parameters,
comparing Eq.~(\ref{zst}) with the $Z$ couplings 
given in Section~\ref{zcurrent} yields
\bea
   \alpha S &\!\!\!=\!\!\!&
       -{2\over Q_E}
         (c^2 - s^2) (g_L^f +g_R^f) \sqrt{L}
         -8 c^2 s^2 
\nonumber
\\
  &&
    + {(g_L^f - g_R^f) \sqrt{L}\over
      T^3_f}
   \left[
     {2\over Q_E}
      (c^2 -s^2) 
       (T^3_f-2 Q_E s^2)
       +8 c^2 s^2 \right] ,
\\
  \alpha T &\!\!\!=\!\!\!&
    2 \left(
      {g_L^f -g_R^f 
      \over T^3_f} -1\right) ,
\eea
where $Q_f \neq 0$ and
the couplings $g_L^f$ and $g_R^f$ 
have been employed as in 
Eq.~(\ref{eq:alr}).
For electron, this evaluation leads to
\bea
   S (\textrm{electron}) = 2.2049 ,\qquad
   T (\textrm{electron}) = 2.72176 .
    \label{formalst}
\eea
The experimental data is
$S (\textrm{Exp.}) = -0.10 \pm 0.10$
and $T(\textrm{Exp.}) = -0.08 \pm 0.11$~\cite{PDG}.
In the experimental constraint, 
loop corrections in the standard model
are taken into account.
The one-loop contribution in the standard model 
is given by
\bea
  S (\textrm{SM 1-loop})
 = 0.247565 ,
\qquad
  T (\textrm{SM 1-loop})
 = 1.25605  , 
\eea
for $m_h =117$GeV. The full equations for the
$S$ and $T$ parameters in the standard model 
are summarized in Appendix~\ref{app:smst}.
The contribution
 (\ref{formalst}) is large compared to 
the loop corrections in the standard model.
We emphasize that Eq.~(\ref{formalst}) is a formal 
equation.
The vacuum polarization in the gauge sector
is expected to be flavor universal.
It might be proper to adopt the picture 
that tree-level corrections are flavor-violated
and correspond to some alternative parameters
except for $S$ and $T$ parameters
and that one-loop vacuum polarization is
the first contribution for
$S$ and $T$ parameters.
In such a case, the experimental constraint should be 
read for the alternative parameters as well as
for the $S$ and $T$ parameters. 
Further investigation will be left to future work.

\section{Summary and discussions \label{conclusion}}

We have presented the tree-level 
prediction of
the forward-backward production asymmetry
on the $Z$ resonance for quarks and leptons,
$A_{FB}^f$,
in the $SO(5)\times U(1)_X$ gauge-Higgs unification
model given in Ref.~\cite{HNU}.
It has been found that 
the tree-level prediction
for $b$ quark production
gives $A_{FB}^b(\textrm{XZZ, FK})=0.09952$,
which is quite close to the central value of 
the experimental data
$A_{FB}^b(\textrm{Exp}.)=0.0992\pm 0.0016$.
We have also found for $c$ quark production  
$A_{FB}^c=0.07073(\textrm{XZZ,FK})$ 
which is also close to the central value of 
the experimental data
$A_{FB}^c(\textrm{Exp}.)=0.0707\pm 0.0035$.
For all fermions, the tree-level predictions of
$A_{FB}^f$ have been given,
and it has been shown that the values are not very sensitive 
to whether masses of quarks and leptons
are taken as running masses or pole masses.
We have also evaluated the $Z$ decay width and 
the $S$ and $T$ parameters.
As for these quantities, it has been shown that
the effect of lepton mixing and the
identification of relevant parameters
are worth examining.

The small deviation from the experimental
data is closely related to the left-right symmetry
similar to the custodial symmetry in the standard model
as shown in Ref.~\cite{HNU},
according to a general discussion in Ref.~\cite{%
Agashe:2006at},
although a numerical analysis is required 
under the present understanding.
The normalized coefficients of mode function for fermions
are non-vanishing only for the part symmetric 
under the exchange of left and right isospin eigenvalues 
in $SU(2)_L\times SU(2)_R$:
$T^{3_L}=T^{3_R} = +{1\over 2}$,
$T^{3_L}= T^{3_R} =-{1\over 2}$,
$T^{3_L}=T^{3_R}=0$ and
$(T^{3_L}=-T^{3_R}=+{1\over 2})\oplus
(T^{3_L}=-T^{3_R}=-{1\over 2})$.
The left-right asymmetric part has the 
coefficients proportional to $\cos\theta_H$
which vanishes at $\theta_H=\pm {1\over 2}\pi$
for the potential minimum.

The scattering process $e^+e^- \to f\bar{f}$ 
receives contributions from 
not only tree level but also quantum loop level.
This must be treated appropriately.
For example,
one-loop corrections to couplings
of heavy quarks to $Z$ boson have 
been shown to be sizable in similar 
models~\cite{Carena:2007ua}.
In the standard model,
radiative effects of heavy fields with masses
much larger than $m_Z$ are dominated by
oblique corrections.
The polarization asymmetry for the decay $Z\to e^+ e^-$ is
corrected as
\bea
   A_{LR}^e  &\!\!\!=\!\!\!&
     {\left[-{1\over 2} + s_*^2 (q^2)\right]^2
       -\left[s_*^2 (q^2)\right]^2
    \over \left[
       -{1\over 2} +s_*^2 (q^2) \right]^2
       +\left[ s_*^2 (q^2)\right]^2} ,
       \label{alrstar}
\eea
and the forward-backward asymmetry for $b$ quark,
for instance, is given by
\bea
  A_{FB}^b &\!\!\!=\!\!\!&
    {3\over 4} 
     A_{LR}^e
     {\left[-{1\over 2} + {1\over 3}s_*^2 (q^2)\right]^2
       -\left[{1\over 3} s_*^2 (q^2)\right]^2
    \over \left[
       -{1\over 2} +{1\over 3}s_*^2 (q^2) \right]^2
       +\left[{1\over 3} s_*^2 (q^2)\right]^2} 
    \left(1-k_A {\alpha_s\over \pi}\right) ,
    \label{afbstar}
\eea
where QCD corrections are included in $k_A$ and the strong 
coupling constant is $\alpha_s$.
In Eqs.~(\ref{alrstar}) and (\ref{afbstar}),
the weak mixing angle is replaced by the effective quantity,
\bea
  s_*^2 (q^2) 
    =\sin^2 \theta_0
   +{\alpha\over c^2 
   -s^2}
      \left({1\over 4}S-
    s^2 c^2 T\right) ,
  \qquad
   \sin 2\theta_0 \equiv
  \left({4\pi \alpha(m_Z)\over \sqrt{2}G_F m_Z^2}
   \right)^{1/2} ,
\eea
where 
the Fermi constant are denoted as 
$G_F$.
When $\alpha(m_Z)$, $G_F$ and $m_Z$ are taken as
input parameters,
the effect of radiative corrections is accommodated in
the $S$ and $T$ parameters 
which is rewritten in the standard model as
\bea
  \alpha S = 
  4e^2 \left[ \Pi_{3_L3_L}'(0) -\Pi_{3_LQ_E}'(0)\right]
  , 
\qquad
  \alpha T =
    {e^2 \over s^2 c^2 m_Z^2}
      \left[\Pi_{1_L1_L}(0) -\Pi_{3_L3_L}(0)\right] ,
\eea
where the vacuum polarizations $\Pi$
have the components for the electric charge $Q_E$ and
the $SU(2)_L$ indices, $1_L,2_L,3_L$.
The $S$ and $T$ parameters in the standard model
are two finite linear combinations 
of vacuum polarizations.
This finiteness is understood 
from a viewpoint that the symmetry of the theory should be recovered
at large momentum, where
$\Pi_{3_L 3_L}|_{\textrm{\scriptsize div}}\sim 
\Pi_{1_L 1_L}|_{\textrm{\scriptsize div}}$
and $(\Pi_{3_L 3_L}|_{\textrm{\scriptsize div}}-
\Pi_{3_L Q_E}|_{\textrm{\scriptsize div}})\sim q^2\textrm{-independent}$. 
In higher-dimensional theory, 
a strategy to extract radiative corrections
would be to describe some observed quantities 
in terms of some other observed quantities as in the standard model.
Then $A_{FB}^f$ should be described in terms of 
physical quantities such as running masses.
Because the gauge-Higgs unification scenario
is based on the gauge principle, 
it may be similar to identify finite combinations 
such as the $S$ and $T$ parameters,
involving recovery of a symmetry at large momentum.
Indeed, finite corrections of $S$ and $T$ parameters
has been given in some extra-dimensional models~\cite{%
Barbieri:2004qk}\cite{%
Barbieri:2003pr}-\cite{Lim:2007ea}.
Particularly,
in an $SO(5)\times U(1)$-invariant formulation,
the $S$ parameter seems to give a value much above the 
current experimental bounds for the Wilson
line phase $\theta_H =\pm \pi/2$~\cite{%
Agashe:2004rs}.
The present model has $SO(4)\times U(1)$ multiplets
on the Planck brane so that
their results for the $S$ parameter
cannot be directly applied to the present model.
In addition, higher-dimensional theory includes couplings 
with negative mass dimension
and a new feature appears differently than
in models with dimensionless parameters and masses.
Even if there is only one interaction 
and usual kinetic energy terms,
radiative corrections in higher-dimensional models lead to 
two point functions with multiple poles~\cite{Uekusa:2009dy}.
This make the treatment of loop corrections complicated.
We leave investigation of these radiative corrections
in the present $SO(5)\times U(1)_X$ model 
to future work.

While we have restricted our attention on 
the forward-backward asymmetry on $Z$ resonance, 
recently a large forward-backward
asymmetry for $t$ quark has been observed~\cite{cdf} 
\bea 
 A_{FB}^t = 0.193 \pm 0.065^{\textrm{\scriptsize stat.}}
   \pm 0.024^{\textrm{\scriptsize syst.}} ,
\eea
at $\sqrt{s}=1.96$TeV.
There seems a discrepancy between this value and  
the standard model prediction. 
In this circumstance, 
there has been a possibility that
the contributions of Kaluza-Klein excitations of gauge bosons
account for the discrepancy
in a warped extra-dimensional model~\cite{%
Djouadi:2009nb}.
Kaluza-Klein particles in the present $SO(5)\times U(1)_X$
model might also yield sizable contributions.

We have found that
the violation of universality at the $Z$ boson
vertices is crucial for obtaining
the results concerning $A_{FB}^f$.
If such a violation is arbitrarily large,
flavor-changing neutral currents would also be large.
In the present model, couplings of photon with
fermions are just the normalization
of the fermion wave functions and 
determined completely by 
the electric charges because 
the photon is a constant mode.
At tree level, the only source of lepton-number
violation is the mixing of the low-mass neutrinos.
Then flavor-changing neutral current processes
such as $\mu\to e\gamma$ are extremely
small unobservable probabilities.
This may also be affected by radiative corrections.
In addition to the issue of the above radiative 
corrections,
the model has to be examined in more detail.

\vspace{8ex}

\subsubsection*{Acknowledgments}

I thank Yutaka Hosotani and Shusaku Noda for
the previous collaboration~\cite{HNU}
on which this research is based.
I am also grateful to Minoru Tanaka and Yoshio Koide 
for helpful suggestions. 
This work is supported by Scientific Grants 
from the Ministry of Education
and Science, Grant No.~20244028.

\newpage

\begin{appendix}

\section{Mode functions of Z boson and fermions \label{app:mode}}

In this appendix, mode functions of 
$Z$ boson and fermions and some related functions
are summarized.

\subsection{Z boson}

The $SO(5)$ gauge fields are
split into classical and quantum parts
$A_M =A_M^c + A_M^q$, where
$A_\mu^c =0$ and $A_y^c = (dz/dy) A_z^c 
=kz A_z^c$. 
With the gauge-fixing functional 
\bea
    f_{\textrm{\scriptsize gf}}^{(A)} = 
    z^2 \left\{ \eta^{\mu\nu} {\cal D}_\mu^c A_\nu^q
    +\xi k^2 z {\cal D}_z^c \left({1\over z} A_z^q\right) \right\} ,
\eea
the quadratic action for the $SO(5)$ gauge fields is
\bea
   S_{\textrm{\scriptsize bulk 2}}^{%
  \textrm{\scriptsize gauge}} =
   \int d^4 x {dz\over kz}
  \left[ \textrm{tr}\left\{
   \eta^{\mu\nu} A_\mu^q (\Box + k^2 {\cal P}_4) A_\nu^q
   +k^2 A_z^q (\Box +k^2 {\cal P}_z)A_z^q 
   \right\} \right] ,
\eea
for $\xi=1$. Here
$A_\mu^c=0$ have been taken.
The differential operators are
$\Box = \eta^{\mu\nu}\partial_\mu \partial_\nu$,
${\cal P}_4= z{\cal D}_z^c ({1/ z}) {\cal D}_z^c$,
${\cal P}_z = {\cal D}_z^c z {\cal D}_z^c ({1/ z})$,
where ${\cal D}_M^c A_N^q =
\partial_M A_N^q - ig_A [ A_M^c, A_N^q]$.   
The linearized equations of motion is
\bea
   \Box A_\mu^q + k^2 z {\cal D}_z^c {1\over z} 
  {\cal D}_z^c A_\mu^q 
   =0 ,
\qquad
   \Box A_z^q + k^2 {\cal D}_z^c z 
  {\cal D}_z^c {1\over z} A_z^q 
   =0 .
 \label{eom1}
\eea

The $SO(4)$ vector $A_y^{\hat{a}}$ which forms
the $SU(2)_L$-doublet
$\Phi_H^t =(
A_y^{\hat{2}} +i A_y^{\hat{1}},
A_y^{\hat{4}} -i A_y^{\hat{3}})$
has zero modes.
One can utilize the residual symmetry such that  
the zero mode of $A_y^{\hat{4}}$ 
yield a nonzero vacuum expectation value 
$\langle A_y^{\hat{a}}\rangle =v\delta^{a4}$.
The Wilson line phase $\theta_H$ is given by
$\exp \{{i\over 2}\theta_H (2\sqrt{2}T^{\hat{4}})\}
=\exp\{ig_A\int_1^{z_L} dz \langle A_z\rangle\}$
so that 
$\theta_H ={1\over 2}
g_A v \sqrt{(z_L^2 -1)/k}$. 
By a large gauge transformation which maintains 
the orbifold boundary conditions, $\theta_H$ can be shifted to
$\theta_H+2\pi$.
The gauge invariance of the theory implies that
physics is periodic in $\theta_H$ with a period $2\pi$.
With a gauge transformation,
a new basis can be taken
in which the background 
field vanishes, $\tilde{A}_z^c=0$.
The new gauge is called the twisted gauge as the boundary
conditions are twisted.
The new gauge potentials are related to the original
ones by$\tilde{A}_M = \Omega A_M^q \Omega^{-1}$,
$\tilde{B}_M =B_M^q$.
Here
$\Omega(z) = \exp \{ i\theta(z) \sqrt{2}T^{\hat{4}}\}$ 
and $\theta(z) =\theta_H(z_L^2 -z^2)/(z_L^2 -1)$.  
The equations of motion (\ref{eom1}) 
become
\bea
     \Box \tilde{A}_\mu + k^2 \left(\partial_z^2 -{1\over z} \partial_z\right) 
   \tilde{A}_\mu =0 ,
\quad
    \Box \tilde{A}_z + k^2 \left(\partial_z^2 -{1\over z}\partial_z +{1\over z^2}\right)
    \tilde{A}_z =0 .
\eea
The field $\tilde{B}_M$
satisfies the same equations as $\tilde{A}_M$.

The four-dimensional components of the
$SO(5)$ and $U(1)_X$ gauge bosons contain
$W$ and $Z$ bosons and photon as
\bea
   \tilde{A}_\mu(x,z)
     &\!\!\!=\!\!\!&
   W_\mu \left\{
      h_W^L T^{-_L} + h_W^R T^{-_R} + h_W^\wedge T^{\hat{-}} \right\}
   +   
    W_\mu^\dag \left\{
      h_W^L T^{+_L} + h_W^R T^{+_R} + h_W^\wedge T^{\hat{+}}
        \right\}
\nonumber
\\
  &&\qquad
   +Z_\mu \left\{
     h_Z^L T^{3_L} +h_Z^R T^{3_R} + h_Z^\wedge T^{\hat{3}} 
  \right\}
  +A^\gamma_\mu
      h_\gamma\left\{ T^{3_L} +T^{3_R}\right\} 
   +\cdots ,
\\
  \tilde{B}_\mu(x,z)&\!\!\!=\!\!\!&
     Z_\mu h_Z^B +A_\mu^\gamma h_\gamma^B 
  +\cdots.
\eea
The wave functions $h_W^L(z)$, $h_W^R(z)$
and $h_W^\wedge(z)$ for $W_\mu(x)$,
$h_Z^L(z)$, $h_Z^R(z)$, $h_Z^\wedge(z)$ and
$h_Z^B(z)$ for $Z(x)$ and
$h_\gamma(z)$ and $h_\gamma^B(z)$ for $A_\gamma(x)$ 
satisfy their own equations of motion and boundary 
conditions. 
For the $Z$ boson tower, for instance,
the boundary conditions at $z=1$ is given as 
\bea
  0&\!\!\!=\!\!\!&
   s_\phi\left(\sin^2{\theta_H\over 2}\,
  \partial_z h_Z^L
  +\cos^2{\theta_H\over 2} \,\partial_z
   h_Z^R
  +{1\over \sqrt{2}} \sin\theta_H \,\partial_z 
  h_Z^\wedge \right)
  +c_\phi \partial_z h_Z^B,
  \label{hzbc1}
\\
  0&\!\!\!=\!\!\!&
  c_\phi
  \left(\sin^2{\theta_H\over 2} \,h_Z^L
  +\cos^2{\theta_H\over 2} \,h_Z^R
  +{1\over \sqrt{2}} \sin\theta_H \,
h_Z^\wedge\right)
  -s_\phi h_Z^B.
    \label{hzbc2}
\eea
The lowest mass modes for 
$W_\mu(x)$, $Z_\mu(x)$ and $A_\gamma(x)$
are $W$ and $Z$ bosons and photon, respectively.

The wave functions for the bosons in the $Z$
boson tower are
\bea
  h_Z^L(z) &\!\!\!=\!\!\!&
    { c_\phi^2 + \cos \theta_H (1+s_\phi^2)
    \over 2\sqrt{1+s_\phi^2}}
       N_Z(z;\lambda) ,
\quad
 h_Z^R(z) =
    { c_\phi^2 - \cos \theta_H (1+s_\phi^2)
     \over 2\sqrt{1+s_\phi^2}}
       N_Z (z;\lambda) ,
\nonumber
\\
  h_Z^\wedge(z)  &\!\!\!=\!\!\!&
     - {\sin\theta_H
       \over \sqrt{2}} \sqrt{1+s_\phi^2}
         D_Z(z;\lambda) ,
\qquad
  h_Z^B(z)  =
    -{s_\phi c_\phi\over \sqrt{1+s_\phi^2}} 
  N_Z(z;\lambda) .
    \label{hz}
\eea
Here $N_Z=2b\sqrt{1+s_\phi^2}C(z;\lambda)$
and 
$D_Z=2b\sqrt{1+s_\phi^2}(C(1;\lambda)/S(1;\lambda))S(z;\lambda)$.
The $C$ and $S$ functions are defined as
\bea
   C(z;\lambda) &\!\!\!=\!\!\!&
      {\pi\over 2}\lambda z z_L F_{1,0}(\lambda z,
        \lambda z_L) ,
\quad
   C'(z;\lambda) =
      {\pi\over 2}\lambda^2 z z_L F_{0,0}(\lambda z,
        \lambda z_L) ,
\\
   S(z;\lambda) &\!\!\!=\!\!\!&
      -{\pi\over 2}\lambda z  F_{1,1}(\lambda z,
        \lambda z_L) ,
\quad
   S'(z;\lambda) =
      -{\pi\over 2}\lambda^2 z F_{0,1}(\lambda z,
        \lambda z_L) .
\eea
where a useful linear combination of Bessel functions is
defined as
\bea
 F_{\alpha,\beta}(u,v)
= J_\alpha (u) Y_\beta(v) 
-Y_\alpha(u) J_\beta(v) ,
\eea 
and
the prime denotes a derivative such as 
$C'=dC/dz$.
A relation 
$CS' -SC' =\lambda z$ holds. 
From the normalization $\int_1^{z_L} dz (kz)^{-1}
\{(h_Z^L)^2 + (h_Z^R)^2 + (h_Z^\wedge)^2
 +(h_Z^B)^2 \} =1$
the coefficient $b$ is determined to be
\bea
  b^{-2}
 &\!\!\!=\!\!\!&
  \int_1^{z_L}
  {dz\over kz}
   \,  2(1+s_\phi^2)^2 
   \left[
    {2\over 1+s_\phi^2}
  (C(z;\lambda))^2
  \right.
\nonumber
\\
  && \left.
  -\sin^2\theta_H
  \left(
   (C(z;\lambda))^2
    -\left({C(1;\lambda)\over S(1;\lambda)}
   \right)^2
  (S(z;\lambda))^2 \right)
\right]   .
\eea   
The equation (\ref{hzbc2}) is automatically
fulfilled with Eq.~(\ref{hz}).

The mass spectrum of the $Z$ tower is determined by 
\bea
   2S(1;\lambda) C'(1;\lambda)
     +\lambda(1+s_\phi^2) \sin^2\theta_H =0 .
     \label{zm}
\eea
The other boundary condition (\ref{hzbc1}) is fulfilled with
Eq.~(\ref{zm}).
The mass of the lightest mode, the $Z$ boson,
is given by
\bea
    m_Z \approx {m_W\over \cos \theta_W} ,
 \qquad
  m_W  \approx {m_\textrm{\scriptsize KK}\over \pi \sqrt{kL}}
  |\sin\theta_H|  ,
\eea
where $m_{\textrm{\scriptsize KK}}\approx
\pi k e^{-kL}$.

\subsection{Fermions}

In terms of the rescaled fields
$\tilde{\Psi}_a = z^{-2} \Omega(z) \Psi_a$
in the twisted gauge where
\bea
  \Omega(z) 
   = \left(\begin{array}{ccc}
      {\bf 1}_3 && \\
         & \cos \theta(z) & \sin \theta(z) \\
         & -\sin\theta(z) & \cos\theta(z) \\
      \end{array}\right) ,
      \qquad
      \theta(z) ={z_L^2-z^2\over 
        z_L^2 -1} \theta_H ,
\eea 
the action for the fermions in the bulk region becomes
\bea
 S_{\textrm{\scriptsize bulk}}^{%
     \textrm{\scriptsize fermion}}
  &\!\!\!=\!\!\!& \sum_a
    \int d^4x {dz\over k} \,
     i\bar{\tilde{\Psi}}_a 
   \bigg\{
      \Gamma^\mu (\partial_\mu
         -ig_A \tilde{A}_\mu -ig_B Q_{Xa}
          \tilde{B}_\mu)
\nonumber
\\
  &&
   +\Gamma^5 \sigma'
   (\partial_z-ig_A \tilde{A}_z -ig_B Q_{Xa}
     \tilde{B}_z)
  -{c^a\over z} \sigma' 
    \bigg\} \tilde{\Psi}_a 
    ,
\eea
If there were no brane interactions, it would obey
\bea
  &&\left\{
     \left( \begin{array}{cc}
        & \sigma \cdot \partial \\
      \bar{\sigma} \cdot \partial & \\
      \end{array}\right)
   -k\left(\begin{array}{cc}
     D_-(c_a) & \\
     & D_+(c_a) \\
   \end{array}\right)\right\}
 \left(\begin{array}{c}
   \tilde{\Psi}_{aR} \\
   \tilde{\Psi}_{aL} \\
   \end{array}\right) =0 ,
\eea
where
$D_{\pm}(c) = \pm (d/dz) +(c/z)$.
Neumann conditions for $\tilde{\Psi}_R$
and $\tilde{\Psi}_L$ are given by
$D_-(c)\tilde{\Psi}_R=0$ and 
$D_+(c)\tilde{\Psi}_L=0$, respectively.
The resulting second order differential equations are
\bea
    \left\{
        \partial^2 - k^2 D_-(c_a) D_+(c_a)
        \right\} \tilde{\Psi}_{aL} =0 ,
\quad
   \left\{
        \partial^2 - k^2 D_+(c_a) D_-(c_a)
        \right\} \tilde{\Psi}_{aR} =0 .
\eea
The basic functions are given by
\bea
   \left(\begin{array}{c}
   C_L \\
   S_L \\
 \end{array}\right) (z;\lambda, c)
 &\!\!\!=\!\!\!&
  \pm {\pi\over 2} \lambda\sqrt{zz_L}
   F_{c+{1\over 2},c\mp{1\over 2}}
 (\lambda z, \lambda z_L) ,
\\
 \left(\begin{array}{c}
   C_R \\
   S_R \\
 \end{array}\right) (z;\lambda, c)
&\!\!\!=\!\!\!&
   \mp{\pi\over 2} \lambda\sqrt{zz_L}
   F_{c-{1\over 2},c\pm {1\over 2}}
 (\lambda z, \lambda z_L)  .
\eea
They satisfy
\bea
   && \left\{
      D_+(c) D_-(c) -\lambda^2 \right\}
      \left(\begin{array}{c}
        C_R (z)\\
        S_R(z) \\
        \end{array}\right) =0 ,
\quad
        \left\{
      D_-(c) D_+(c) -\lambda^2 \right\}
      \left(\begin{array}{c}
        C_L (z)\\
        S_L(z) \\
        \end{array}\right) =0 ,
\nonumber
\\
  && S_L(z;\lambda ,-c) =-S_R(z;\lambda,c) ,
 \qquad
C_LC_R -S_L S_R=1 .
\eea
They satisfy the boundary conditions that
$C_R=C_L=1$,
$D_- C_R=D_+ C_L =0$,
$S_R=S_L=0$ and
$D_- S_R =D_+ S_L =\lambda$ at $z=z_L$.
Further $D_\pm$ links $L$ and $R$ functions by
$D_+(C_L,S_L) =\lambda(S_R, C_R)$
and $D_-(C_R,S_R) =\lambda(S_L, C_L)$.

\subsubsection*{Top quark}

In the quark sector we chose $c_1=c_2=c$.
The top quark 
component $t(x)$ in four dimensions is contained in the form
\bea
  \left(\begin{array}{c}
  \tilde{U}_L \\
  (\tilde{B}_L \pm \tilde{t}_L)/\sqrt{2} \\
   \tilde{t}_L' \\
  \end{array}\right) (x,z)
   &\!\!\!=\!\!\!&
  \sqrt{k} \left(\begin{array}{c}
  a_U C_L(z;\lambda, c) \\
   a_{B\pm t} C_L(z; \lambda, c) \\
   a_{t'} S_L(z;\lambda, c) \\
  \end{array}\right) t_L(x) ,
\\
   \left(\begin{array}{c}
  \tilde{U}_R \\
  (\tilde{B}_R \pm \tilde{t}_R)/\sqrt{2} \\
   \tilde{t}_R' \\
  \end{array}\right) (x,z)
   &\!\!\!=\!\!\!&
  \sqrt{k} \left(\begin{array}{c}
  a_U S_R(z;\lambda, c) \\
   a_{B\pm t} S_R(z; \lambda, c) \\
   a_{t'} C_R(z;\lambda, c) \\
  \end{array}\right) t_R(x) ,
\eea
where $u$ and $c$ quarks are described in a similar way.
We suppose that the scale of brane 
masses in much larger than the Kaluza-Klein scale;
$\mu_1^2, \mu_2^2 ,\mu_3^2, \tilde{\mu}^2 
\gg m_{\textrm{\scriptsize KK}}$.
Then the ratios of the coefficients $a$ are given by
\bea
   [ a_U , a_{B-t} , a_{t'} ]
   \simeq 
   \left[ -{\sqrt{2}\tilde{\mu}
    \over \mu_2} ,
    -c_H, 
    -{s_H C_L|_{z=1}\over 
  S_L|_{z=1} }
  \right]
    a_{B+t} .
\eea    
Here $c_H =\cos\theta_H$ and $s_H =\sin\theta_H$. 
The coefficient $a_{B+t}$ is determined
by
\bea
  a_{B+t}^{-1}
    =\int_1^{z_L}
     dz
    \left\{
    \left(2\left({\tilde{\mu}\over \mu_2}
      \right)^2 +1
      +c_H^2\right)
      (C_L(z))^2
    +s_H^2
      \left({C_L|_{z=1}
       \over S_L|_{z=1}}\right)^2
     (S_L(z))^2 
     \right\} .
\eea
The top quark mass $m_t=k\lambda_t$ obeys
\bea
  \tilde{\mu}^2 S_R
    C_L + \mu_2^2 
     C_L \left\{
  S_R + {s_H^2\over 2S_L}\right\}
      \bigg|_{z=1,\lambda=\lambda_t} =0 .
      \label{tmass}
\eea
This equation
includes the ratio 
$\tilde{\mu}/\mu_2$
and $c$ as parameters.
There is
the corresponding equation for bottom quark mass $m_b$
which includes the same parameters.

\subsubsection*{Bottom quark}

The bottom quark component $b(x)$ is contained in the form
\bea
   \left(\begin{array}{c}
  \tilde{b}_L \\
  {1\over \sqrt{2}} (\tilde{D}_L \pm \tilde{X}_L) \\
   \tilde{b}_L' \\
  \end{array}\right) (x,z)
   &\!\!\!=\!\!\!&
  \sqrt{k} \left(\begin{array}{c}
  a_b C_L(z;\lambda, c_1) \\
   a_{D\pm X} C_L(z; \lambda, c_2) \\
   a_{b'} S_L(z;\lambda, c_2) \\
  \end{array}\right) b_L(x) ,
\\
   \left(\begin{array}{c}
  \tilde{b}_R \\
  {1\over \sqrt{2}} (\tilde{D}_R \pm \tilde{X}_R) \\
   \tilde{b}_R' \\
  \end{array}\right) (x,z)
   &\!\!\!=\!\!\!&
  \sqrt{k} \left(\begin{array}{c}
  a_b S_R(z;\lambda, c_1) \\
   a_{D\pm X} S_R(z; \lambda, c_2) \\
   a_{b'} C_R(z;\lambda, c_2) \\
  \end{array}\right) b_R(x) ~.
\eea
The $d$ and $s$ quarks are described in a similar manner.
The ratios of the coefficients are given by
\bea
 [ a_b , a_{D-X}, a_{b'} ]
  =\left[
    -{\sqrt{2} \mu_2 \over \tilde{\mu}} ,
    c_H,
    {s_H C_L |_{z=1}\over S_L |_{z=1}} \right] a_{D+X} ~.
\eea
The coefficient $a_{D+X}$ is given by
\bea
  a_{D+X}^{-2}
    =\int_1^{z_L}
     dz
    \left\{
    \left(2\left({\mu_2 \over \tilde{\mu}}
      \right)^2 +1
      +c_H^2\right)
      (C_L(z))^2
    +s_H^2
      \left({C_L|_{z=1}
       \over S_L|_{z=1}}\right)^2
     (S_L(z))^2 
     \right\} .
\eea
The mass $m_b = k \lambda_b$   obeys
\bea
  \mu_2^2 S_R
    C_L + \tilde{\mu}^2 
     C_L \left\{
  S_R + {s_H^2\over 2S_L}\right\}
      \bigg|_{z=1, \lambda=\lambda_b} =0 .
      \label{bmass}
\eea

Combining Eqs.\ (\ref{tmass}) and (\ref{bmass}), one finds
\bea
 \frac{\tilde\mu^2}{\mu_2^2} 
&\!\!\!=\!\!\!&
 - \bigg\{ 1 + \frac{s_H^2}{2 S_L(1; \lambda_t, c) 
  S_R(1; \lambda_t, c)} \bigg\} 
 \nonumber
 \\
&\!\!\!=\!\!\!&
 -  \bigg\{ 1 + \frac{s_H^2}{2 S_L(1; \lambda_b, c) 
    S_R(1; \lambda_b, c)} \bigg\}^{-1}~.
   \label{muratio}
\eea
The value of $\theta_H$ is dynamically determined. 
In the present model 
$\theta_H = \pm {1\over 2}\pi$.
Hence, given $k$, $m_t$, $m_b$, the parameters
$c$ and $|\tilde\mu/\mu_2|$ are determined. 
These values are not very sensitive 
on the value of $k$.

The wave functions
$C_L$ and $S_L$ for 
the left-handed quarks $t_L$, $b_L$
are localized near the Planck brane, whereas
$C_R$ and $S_R$ for the right-handed
quarks $t_R$, $b_R$
are localized near the TeV brane.

\subsubsection*{Tau lepton and tau neutrino}

We need to introduce only a vector multiplet $\Psi_3$ in 
Eqs.~(\ref{psi3}) and (\ref{psi4})
 to describe a massless neutrino for each generation.
The $\tau$ lepton is  contained in the form
\bea
   \left(\begin{array}{c}
  {1\over \sqrt{2}} (\tilde{\tau}_L \pm \tilde{L}_{1XL}) \\
   \tilde{\tau}_L' \\
  \end{array}\right) (x,z)
   &\!\!\!=\!\!\!&
  \sqrt{k} \left(\begin{array}{c}
   a_{\tau \pm L_{1X}} C_L(z; \lambda, c_3) \\
   a_{\tau'} S_L(z;\lambda, c_3) \\
  \end{array}\right) \tau_L(x) ,
\nonumber
\\  
   \left(\begin{array}{c}
  {1\over \sqrt{2}} (\tilde{\tau}_R \pm \tilde{L}_{1XR}) \\
   \tilde{\tau}_R' \\
  \end{array}\right) (x,z)
   &\!\!\!=\!\!\!&
  \sqrt{k} \left(\begin{array}{c}
   a_{\tau \pm L_{1X}} S_R(z; \lambda, c_3) \\
   a_{\tau'} C_R(z;\lambda, c_3) \\
  \end{array}\right) \tau_R(x) .
\eea
The $e$ and $\mu$ leptons are described in a similar way.
The ratios of the coefficients are given by
\bea
 [ a_{\tau-L_{1X}}, a_{\tau'} ]
  \simeq \left[
    c_H,
    {s_H C_L (z=1)\over S_L (z=1)} \right] 
 a_{\tau +L_{1X}},
\eea
for $(\mu_1^\ell)^2 \gg m_{\textrm{\scriptsize KK}}$.
The coefficient $a_{\tau +L_{1X}}$ is given by
\bea
  a_{\tau +L_{1X}}^{-2}
    =\int_1^{z_L}
     dz
    \left\{
    \left(1
      +c_H^2\right)
      (C_L(z))^2
    +s_H^2
      \left({C_L(z=1)
       \over S_L(z=1)}\right)^2
     (S_L(z))^2 
     \right\} .
\eea
The mass $m_\tau = k \lambda_\tau$  is determined by
\bea
   (\mu_1^\ell )^2 
     C_L \left\{
  S_R + {s_H^2\over 2S_L}\right\}
      \bigg|_{z=1} =0 .
      \label{leptonmass1}
\eea
The $\nu_\tau$ neutrino is contained in the form
\bea
  \tilde{\nu}_{\tau L} (x,z)
    =\sqrt{k} a_{\nu_\tau} C_L(z;0,c) \nu_{\tau L}(x) .
\eea
The $\nu_e$ and $\nu_\mu$ neutrinos are described
in a similar way.
For massless fermions, the $C$ function
becomes $C_L(z;0,c)  =(z_L/z)^c$. 
The coefficient is written as a simple equation
$a_{\nu_\tau} =\sqrt{(2c-1)/(z_L^{2c}-z_L)}$.

To describe massive neutrinos one needs to introduce two multiplets
$\Psi_3$ and $\Psi_4$.  The structure is the same as 
in the quark sector.
We choose $c_3=c_4=c$.
Equations in the lepton sector 
are obtained with
the correspondence between leptons and quarks:
\bea
 (\nu_\tau,L_{2Y}, L_{3X}, \nu'_\tau)
  &\!\!\!\leftrightarrow\!\!\!&
(U, B, t, t'),
  \qquad
(\hat{L}_{3XR}, \hat{L}_{2YR})
  \leftrightarrow
(\hat{U}_R, \hat{B}_R),
\nonumber
\\
(L_{3YL},\tau,L_{1X}, \tau')
 &\!\!\!\leftrightarrow\!\!\!&
(b, D, X, b') ,
\qquad
(\hat{L}_{3YR}, \hat{L}_{1XR})
 \leftrightarrow
(\hat{D}_R, \hat{X}_R) ,
\nonumber
\\
(\mu_1^\ell, \mu_2^\ell, \mu_3^\ell, \tilde{\mu}^\ell)
 &\!\!\!\leftrightarrow\!\!\!&
(\mu_3, \mu_1, \tilde{\mu}, \mu_2) .
\eea

The behavior of localization for wave functions
in the lepton sector
is similar to that in the quark sector.

\section{
The asymmetries for a large warp factor
\label{app:wi}}

In this appendix, we take
$z_L =1.0 \times 10^{17}$
and $k=5.0\times 10^{19}$GeV
as input parameters for warped geometry.
For these values of $z_L$ and $k$, the Kaluza-Klein scale 
is given by $m_{\textrm{\scriptsize KK}}=1.57$TeV.
For the set XZZ, the polarization asymmetry 
$A_{LR}^f$ and 
the forward-backward asymmetry $A_{FB}^f$ are shown
in Table~\ref{tab:wi}.
\begin{table}[htb]
\begin{center}
\caption{$A_{LR}^f$ and $A_{FB}^f$ for the set XZZ where
$z_L=1.0\times 10^{17}$ and $k=5.0\times 10^{19}$GeV. 
The deviation between the prediction and the experimental
value is denoted in parentheses. 
\label{tab:wi}}
\vskip 10pt
\begin{tabular}{cccc}
\hline\hline
$f$ & $A_{LR}^f$ & $A_{FB}^f$ & Pull \\
\hline
$u$ &
0.6649  & 
0.07124 & 
\\
$c$ &
0.6650  &  
0.07125 &  
-0.2       
\\
$t$ &
0.5310  & 
0.05690 & 
\\
\hline
$d$ &
0.9349  & 
0.10017 & 
\\
$s$ &
0.9349  & 
0.10018 & 
-0.2      
\\
$b$ &
0.9351  & 
0.10019 & 
-0.6      
\\ \hline
$e$ &
0.1429  & 
0.01531 & 
-0.3      
\\
$\mu$ &
0.1431  & 
0.01533 & 
1.2       
\\     
$\tau$ &
0.1432  & 
0.01534 & 
2.0       
\\
\hline
\end{tabular}
\end{center}
\end{table}
For this large warp factor,
the tree-level predictions of the model 
are also
close to the central values
of the experimental data.

\section{The values 
of $S$ and $T$ parameters
at one-loop in the standard model.
\label{app:smst}}

In this appendix, the equations for 
one-loop corrections for the $S$ and $T$ parameters
in the standard model are summarized~\cite{%
Marciano:1980pb}-%
\cite{PeskinSchroeder}.

Each vacuum polarization diagram includes
logarithmic divergence 
\bea
   E={2\over \epsilon} -\gamma + \log
  \left({4\pi\over M^2}\right) .
\eea
The other part can be described with
Feynman parameter integrals
\bea
   b_0(12 X) &\!\!\!=\!\!\!&
     \int_0^1 dx \, \log
        \left({\Delta(M_1^2, M_2^2 , q_X^2) \over M^2}
    \right) ,
\\
  b_1(12 X) & \!\!\!=\!\!\!&
    \int_0^1 dx \, x\log 
 \left({\Delta(M_1^2, M_2^2,
       q_X^2)\over M^2}\right) ,
\\
  b_2(12 X) & \!\!\!=\!\!\!&
    \int_0^1 dx \, x(1-x)\log \left(
   {\Delta(M_1^2, M_2^2,
       q_X^2)\over M^2}\right) ,
\eea
where $q_q^2 \equiv  q^2$ and
$\Delta (M_1^2 , M_2^2 ,q^2) 
= x M_2^2 +(1-x) M_1^2 +x(1-x)q^2$.

The top and bottom loops contribute
\bea
 S_{tb} &\!\!\!=\!\!\!& -{1\over 2\pi}
    \left(1+{1\over 3} \log {m_b^2 \over m_t^2}\right) ,
\\
  T_{tb}
 &\!\!\!=\!\!\!& {3\over 8\pi s^2 c^2 m_Z^2 }
      \left( {m_t^2 m_b^2 \over m_t^2 -m_b^2}
      \log {m_b^2\over m_t^2}
    +{m_t^2 +m_b^2\over 2}\right) .
\eea
The Higgs boson loops contribute
\bea
 S_h &\!\!\!=\!\!\!&
      {1\over \pi}
   \left\{
   E\left(-{1\over 12}\right)
   -{1\over 4}(m_Z^2 -m_h^2) [2 b_1'(h Z 0) 
  -b_0'(h Z 0)]
  \right.
\nonumber
\\
   && \left.
   -{1\over 4}[4 b_2(h Z 0)-b_0(h Z 0)] 
  -m_Z^2 b_0'(h Z 0)
  \right\} ,
\\
  T_h &\!\!\!=\!\!\!&
   {1\over 4\pi s^2 c^2 m_Z^2}
  \left\{
       E(m_W^2 -m_Z^2)
     -{1\over 4}\left[
       {3 m_W^2 m_h^2\over m_W^2 -m_h^2}
       \log{m_W^2\over m_h^2}
     +4 m_W^2 \log {m_W^2\over M^2} 
      \right. \right.
\nonumber
\\
   && \left.\left.
  -{3 m_Z^2 m_h^2\over m_Z^2 -m_h^2}
       \log{m_Z^2\over m_h^2}
     - 4 m_Z^2 \log {m_Z^2\over M^2}
  -{7\over 2} (m_W^2 -m_Z^2) \right]\right\} ,
\eea
which depend on $E$ and $M$.
The sum of $S_h$ and the contribution to $S$
from gauge boson loops is independent of $E$ and $M$.
Also 
the sum of $T_h$ and the contribution to $T$
from gauge boson loops is independent of $E$ and $M$.
Here
\bea
   b_0'(h Z 0)
   &\!\!\!=\!\!\!&
     {1\over m_Z^2 -m_h^2}
     \left[-{1\over 2}
       +{m_Z^2\over m_Z^2 -m_h^2}
        -{m_Z^2 m_h^2\over (m_Z^2 -m_h^2)^2}
        \log {m_Z^2\over m_h^2} \right] ,
\nonumber
\\
  b_1'(h Z 0)
    &\!\!\!=\!\!\!&
      {1\over m_Z^2 -m_h^2}
        \left[ -{1\over 3}
        +{m_Z^2\over 2(m_Z^2 - m_h^2)}
      -{m_Z^2 m_h^2\over (m_Z^2 -m_h^2)^2}
      +{m_Z^2 m_h^4\over 
         (m_Z^2-m_h^2)^3}
          \log {m_Z^2\over m_h^2} \right]
\nonumber
\\
 b_0(h Z 0) &\!\!\!=\!\!\!&
     -1 + \log {m_h^2\over M^2}
   + {m_Z^2\over m_Z^2-m_h^2}\log {m_Z^2\over m_h^2} ,
\nonumber
\\
  b_2(h Z 0) &\!\!\!=\!\!\!&
   {1\over 6}\log{m_Z^2\over M^2}
   -{5\over 36}
  +{m_h^2\over 3(m_Z^2 -m_h^2)}
   -{m_h^4\over 2(m_Z^2 -m_h^2)^2}
   \log{m_Z^2\over m_h^2}
\nonumber
\\
  &&
  + {m_h^4\over 3(m_Z^2-m_h^2)^2}
  -{m_h^6\over 3(m_Z^2 - m_h^2)^3}
  \log{m_Z^2\over m_h^2}.
\eea
The gauge boson loops contribute
\bea
  S_g &\!\!\!=\!\!\!&
 {1\over \pi}
    \left\{
      E\left({1\over 12}\right)
       +(b_2 -{1\over 4} b_0)(WW 0) \right\}
  ,
\\  
   T_g &\!\!\!=\!\!\!&
   {1\over 4\pi s^2 c^2 m_z^2}
 \left\{
  E(m_Z^2 -m_W^2 )
    \right.
   +(2 c^2 + {1\over 4})    
       (m_W^2 -m_Z^2) (2 b_1- b_0) (W Z 0)
\nonumber
\\
  &&
  \left. -(m_Z^2 -3 m_W^2) b_0(W Z 0)
      -2 m_W^2 b_0 (WW 0) 
    +2s^2 m_W^2 (2 b_1 -b_0) (W 00) \right\}  .
\eea
Here
\bea
  b_0(WW 0) &\!\!\!=\!\!\!&
    \log {m_W^2 \over M^2} ,
\\
  b_2 (WW 0) &\!\!\!=\!\!\!&
      {1\over 6} \log {m_W^2 \over M^2} ,
\\
   b_0 (W Z 0) 
   &\!\!\!=\!\!\!&
       \log {m_Z^2 \over M^2}
          -1 + 
        {m_W^2 \over m_Z^2 -m_W^2} 
          \log {m_Z^2 \over m_W^2} ,
\\
   b_1 (W Z 0) &\!\!\!=\!\!\!&
       {1\over 2}\log {m_Z^2 \over M^2}
          -{1\over 4} +{1\over 2} 
           {m_W^2 \over m_Z^2 -m_W^2}
     -{1\over 2}
       {m_W^4 \over (m_Z^2 -m_W^2)^2}
        \log {m_Z^2 \over m_W^2}
         ,
\\
   b_0 (W 00 )
   &\!\!\!=\!\!\!&
    \log {m_W^2 \over M^2} -1 ,
\\
   b_1 (W 00)
  &\!\!\!=\!\!\!&
    {1\over 2}\log {m_W^2 \over M^2} -{3\over 4} .
\eea

For the input values given in Section~\ref{afb},
the $S$ and $T$ parameters become 
\bea
 && S_{tb} =  0.2742 , 
\qquad  T_{tb} = 1.1853 , 
\\
 && S_h + S_g = -0.02666 , 
  \qquad
   T_h + T_g = 0.07076 , 
   \qquad \textrm{for~} m_h= 117\textrm{GeV},
\\   
  && S_h + S_g = 0.04866 , 
  \qquad
   T_h + T_g = 0.1701 , 
   \qquad \textrm{for~} m_h= 340 \textrm{GeV},
\\
 && S_h + S_g = 0.1108 , 
  \qquad
   T_h + T_g = 0.3121 , 
   \qquad \textrm{for~} m_h= 1000\textrm{GeV} ,
\eea
where the Higgs boson mass in the standard model
has been taken as
an input parameter.
The total contributions to
the $S$ and $T$ parameters,
$S_{tb} +S_h +S_g$ and $T_{tb} +T_h + T_g$, are 
\bea
 S \textrm{(SM 1-loop)}
 &\!\!\!=\!\!\!& 0.2476 , 
\qquad
  T \textrm{(SM 1-loop)}
  = 1.2561 , 
  \qquad \textrm{for~} m_h =117 \textrm{GeV} ,
\\
  S \textrm{(SM 1-loop)}
 &\!\!\!=\!\!\!& 0.3229 , 
\qquad
  T  \textrm{(SM 1-loop)}
  = 1.3554 , 
  \qquad \textrm{for~} m_h =340 \textrm{GeV} ,
\\
  S \textrm{(SM 1-loop)}
 &\!\!\!=\!\!\!& 0.3850 , 
\qquad
  T \textrm{(SM 1-loop)}
  = 1.4974 , 
  \qquad \textrm{for~} m_h =1000 \textrm{GeV} .
\eea

\end{appendix}

\newpage



\end{document}